\documentclass[sigconf, screen]{acmart}

\renewcommand\footnotetextcopyrightpermission[1]{}

\setcopyright{none}
\AtBeginDocument{%
  \providecommand\BibTeX{{%
    Bib\TeX}}}

\newif\ifcomments
\newif\ifchanges
\commentsfalse\changesfalse

\usepackage{xspace}
\usepackage[capitalise]{cleveref}

\usepackage{soul}
\usepackage{thmtools}
\usepackage{thm-restate}

\usepackage[utf8]{inputenc}
\usepackage{graphicx}
\usepackage{amsmath}
\usepackage{amsthm}
\usepackage{longtable}
\usepackage{booktabs}
\usepackage{algorithm}
\usepackage{algorithmic}
\usepackage{adjustbox}
\usepackage{makecell}
\usepackage{listings}
\usepackage{tikz}
\usepackage{array}
\usetikzlibrary{arrows.meta}
\usetikzlibrary{automata,positioning}
\usetikzlibrary{calc}
\usepackage{tabularx}
\RequirePackage{tikz}

\newcommand{\N}{\ensuremath{\mathbb{N}}}

\newcommand{\Q}{\ensuremath{\mathbb{Q}}}
\newcommand{\R}{\ensuremath{\mathbb{R}}}

\newcommand{\bigO}{\ensuremath{\mathcal{O}}}

\newcommand{\df}{\ensuremath{\mathrel{\smash{\stackrel{\scriptscriptstyle{
    \text{def}}}{=}}}} \;}

\newcommand  {\myclass} [1]  {\ensuremath{\textsf{\upshape #1}}}

\newcommand{\StaClass}[1]{\myclass{#1}\xspace}

\newcommand  {\algorithmicProblem} [1] {\normalfont{\textsc{#1}}\xspace}
\newcommand{\columnWidth}{10cm}
\newcommand{\problemIndent}{\hspace{5mm}}
\newcommand{\algorithmicProblemDescription}[4][10cm]{
    \def\Name{#2}
    \def\Input{#3}
    \def\Question{#4}
      \setlength{\tabcolsep}{1mm}
      \begin{tabular}{rp{#1}r}%
      \textit{Problem:}&\algorithmicProblem{\Name} \\
     \textit{Input:}&\Input \\
     \textit{Question:}&\Question
     \end{tabular}%
    }

\newcommand  {\querydescr} [3] {
\vspace{3mm}
\def\Name{#1}
\def\Input{#2}
\def\Question{#3}
  \problemIndent\begin{tabular}{r p{\columnWidth}r}%
  \textit{Query:} & \algorithmicProblem{\Name} \\
  \textit{Input:} & \Input \\
  \textit{Question:} & \Question
  \end{tabular}
\vspace{3mm}
}

\newcommand     {\LogCFL}     {\StaClass{LOGCFL}}
\newcommand     {\LOGCFL}     {\StaClass{LOGCFL}}
\newcommand     {\LOGSPACE}     {\StaClass{LogSpace}}
\newcommand     {\NL}   {\StaClass{NL}}
\renewcommand   {\P}    {\StaClass{P}}
\newcommand     {\PTIME}    {\myclass{PTime}}
\newcommand     {\NP}   {\StaClass{NP}}
\newcommand     {\ExpTime}  {\StaClass{EXPTIME}}
\newcommand     {\NC}   {\StaClass{NC}}
\newcommand{\NCo}{\mbox{\myclass{NC}$^1$}\xspace}
\newcommand     {\AC}   {\StaClass{AC}}
\newcommand     {\TC}   {\myclass{TC}}
\newcommand     {\ACC}   {\myclass{ACC}}
\newcommand{\ACz}{\mbox{\myclass{AC}$^0$}\xspace}
\newcommand{\TCz}{\mbox{\myclass{TC}$^0$}\xspace}
\newcommand{\ACzmtwo}{\mbox{\myclass{AC}$^0[2]$}\xspace}
\newcommand{\ACo}{\mbox{\myclass{AC}$^1$}\xspace}
\newcommand{\qDynAC}{\ensuremath{\text{q-}\DynClass{AC}}}
\newcommand     {\Sym}   {\myclass{Sym}^+}

\newcommand{\quant}{\mathbb{Q}}
\newcommand{\FO}{\StaClass{FO}}
\newcommand{\FOparity}{\StaClass{FO+Parity}}
\newcommand{\FOar}{\StaClass{FO$(\leq,+,\times)$}}%
\newcommand{\MSO}{\StaClass{MSO}}
\newcommand{\SO}{\StaClass{SO}}
\newcommand{\SOleq}{\StaClass{SO$(\leq)$}}
\newcommand{\SOar}{\StaClass{SO$(\leq,+,\times)$}}
\newcommand{\GSO}{\StaClass{GSO}}
\newcommand{\EMSO}{\StaClass{$\exists$MSO}}
\newcommand{\QFO}[1][\quant]{\StaClass{\ensuremath{#1}FO}}
\newcommand{\EFO}{\QFO[\exists^*]}
\newcommand{\AFO}{\QFO[\forall^*]}
\newcommand{\AEFO}{\StaClass{$\forall/\exists$FO}}
\newcommand{\CQ}[1][]{\StaClass{CQ}}
\newcommand{\UCQ}[1][]{\StaClass{UCQ}}
\newcommand{\CQneg}[1][]{\StaClass{CQ\ensuremath{^{\mneg}}}}
\newcommand{\UCQneg}[1][]{\StaClass{UCQ\ensuremath{^{\mneg}}}}
\newcommand{\Prop}{\StaClass{Prop}}
\theoremstyle{plain}
\newtheorem{innercustomthm}{Theorem}

\newtheorem{innercustomcor}{Corollary}

\theoremstyle{definition}
\newtheorem*{question*}{Question}
\newtheorem*{openquestion*}{Open question}

\providecommand {\calA}      {{\mathcal A}\xspace}
\providecommand {\calB}      {{\mathcal B}\xspace}
\providecommand {\calC}      {{\mathcal C}\xspace}
\providecommand {\calD}      {{\mathcal D}\xspace}
\providecommand {\calE}      {{\mathcal E}\xspace}
\providecommand {\calF}      {{\mathcal F}\xspace}
\providecommand {\calG}      {{\mathcal G}\xspace}
\providecommand {\calH}      {{\mathcal H}\xspace}
\providecommand {\calK}      {{\mathcal K}\xspace}
\providecommand {\calI}      {{\mathcal I}\xspace}
\providecommand {\calL}      {{\mathcal L}\xspace}
\providecommand {\calM}      {{\mathcal M}\xspace}
\providecommand {\calN}      {{\mathcal N}\xspace}
\providecommand {\calO}      {{\mathcal O}\xspace}
\providecommand {\calP}      {{\mathcal P}\xspace}
\providecommand {\calO}      {{\mathcal O}\xspace}
\providecommand {\calQ}      {{\mathcal Q}\xspace}
\providecommand {\calR}      {{\mathcal R}\xspace}
\providecommand {\calS}      {{\mathcal S}\xspace}
\providecommand {\calT}      {{\mathcal T}\xspace}
\providecommand {\calU}      {{\mathcal U}\xspace}
\providecommand {\calV}      {{\mathcal V}\xspace}
\providecommand {\calX}      {{\mathcal X}\xspace}
\providecommand {\calY}      {{\mathcal Y}\xspace}
\providecommand {\calZ}      {{\mathcal Z}\xspace}

\ifcomments
\newcommand{\commentbox}[1]{\noindent\framebox{\parbox{0.98\linewidth}{#1}}}
\newcommand{\todo}[1]{\ \\ {\color{red} \fbox{\parbox{0.98\linewidth}{{\sc
          ToDo}:\\  #1}}}}

\newcommand{\acomment}[2]{\ \\ \fbox{\parbox{0.98\linewidth}{{\sc #1}: #2}}}
\newcommand{\mcomment}[2]{{\color{blue}(#1)}\footnote{#1: #2}} %
\else
\newcommand{\commentbox}[1]{}
\newcommand{\mcomment}[2]{}
\newcommand{\acomment}[2]{}
\fi

\ifchanges

\newcommand{\loldnew}[3]{\commentbox{{\textcolor{blue}{\setlength{\fboxsep}{1pt}\fbox{\small
          #1}}} \textcolor{red}{\footnotesize #2}}
  \textcolor{blue}{#3}}
\setul{}{0.2mm}
\setstcolor{red}
\newcommand{\oldnew}[3]{{\textcolor{blue}{\setlength{\fboxsep}{1pt}\fbox{\small
        #1}}} \st{\footnotesize #2} {\color{blue}#3}}%

\else
\newcommand{\loldnew}[3]{#3}
\newcommand{\oldnew}[3]{#3}
\fi

\newcommand{\tzm}[1]{\mcomment{TZ}{#1}}
\newcommand{\fvm}[1]{\mcomment{FV}{#1}}
\newcommand{\msm}[1]{\mcomment{MS}{#1}}
\newcommand{\tkm}[1]{\mcomment{TK}{#1}}

\newcommand{\tz}[1]{\acomment{TZ}{#1}}
\newcommand{\fv}[1]{\acomment{FV}{#1}}  
\newcommand{\ms}[1]{\acomment{MS}{#1}}
\newcommand{\tk}[1]{\acomment{TK}{#1}}

\newcommand{\tzon}[2][]{\oldnew{TZ}{#1}{#2}} 
\newcommand{\fvon}[2][]{\oldnew{FV}{#1}{#2}}
\newcommand{\mson}[2][]{\oldnew{MS}{#1}{#2}} 
\newcommand{\tkon}[2][]{\oldnew{TK}{#1}{#2}}

\newcommand{\tzlon}[2][]{\loldnew{TZ}{#1}{#2}} 
\newcommand{\fvlon}[2][]{\loldnew{FV}{#1}{#2}} 
\newcommand{\mslon}[2][]{\loldnew{MS}{#1}{#2}} 
\newcommand{\tklon}[2][]{\loldnew{TK}{#1}{#2}} 
\newcommand{\tu}[1]{\langle #1 \rangle}

\RequirePackage{adjustbox}
\newcommand{\toplabelled}[2]{%
\setlength{\tabcolsep}{1pt}
\begin{tabular}{r l}
    \adjustbox{valign=t}{#1:} &
    \adjustbox{valign=t}{#2}
\end{tabular}%
}

\newcommand{\midlabelled}[2]{%
\setlength{\tabcolsep}{1pt}
\begin{tabular}{r l}
    \adjustbox{valign=m}{#1:} &
    \adjustbox{valign=m}{#2}
\end{tabular}%
}
\RequirePackage{tikz}
\tikzstyle{narrownode} = [draw, circle, inner sep=1pt]
\tikzstyle{narrowcoverednode} = [circle,fill=orange!40, inner sep=1pt]
\definecolor{solutionCol}{HTML}{f49016}
\tikzstyle{solutionNode} = [draw = solutionCol, very thick]
\tikzstyle{variableSubtree} = [draw, shape=regular polygon, regular polygon sides=5, inner sep=-3pt, rounded corners]

\usetikzlibrary{positioning}
\usetikzlibrary{arrows,backgrounds,calc}
\usetikzlibrary{decorations.shapes}
\pgfdeclarelayer{background}
\pgfsetlayers{background,main}
\usetikzlibrary{
  arrows.meta,
  calc,
  shapes.geometric,
  decorations.pathreplacing,
  decorations.pathmorphing,
  positioning
}

\newcommand{\convexpath}[2]{
[   
    create hullnodes/.code={
        \global\edef\namelist{#1}
        \foreach [count=\counter] \nodename in \namelist {
            \global\edef\numberofnodes{\counter}
            \node at (\nodename) [draw=none,name=hullnode\counter] {};
        }
        \node at (hullnode\numberofnodes) [name=hullnode0,draw=none] {};
        \pgfmathtruncatemacro\lastnumber{\numberofnodes+1}
        \node at (hullnode1) [name=hullnode\lastnumber,draw=none] {};
    },
    create hullnodes
]
($(hullnode1)!#2!-90:(hullnode0)$)
\foreach [
    evaluate=\currentnode as \previousnode using \currentnode-1,
    evaluate=\currentnode as \nextnode using \currentnode+1
    ] \currentnode in {1,...,\numberofnodes} {
  let
    \p1 = ($(hullnode\currentnode)!#2!-90:(hullnode\previousnode)$),
    \p2 = ($(hullnode\currentnode)!#2!90:(hullnode\nextnode)$),
    \p3 = ($(\p1) - (hullnode\currentnode)$),
    \n1 = {atan2(\y3,\x3)},
    \p4 = ($(\p2) - (hullnode\currentnode)$),
    \n2 = {atan2(\y4,\x4)},
    \n{delta} = {-Mod(\n1-\n2,360)}
  in 
    {-- (\p1) arc[start angle=\n1, delta angle=\n{delta}, radius=#2] -- (\p2)}
}
-- cycle
}

\tikzset{dotted pattern/.style args={#1 and #2}{
   decorate,
   fill,
   decoration={
    shape backgrounds,
    shape=circle,
    shape size=#1,
    shape sep={#2, between center}, 
    },
    draw=none
  },
  dotted pattern/.default={1pt and 1.5mm},
}
\RequirePackage{xifthen}
\RequirePackage{adjustbox}
\RequirePackage{tikz}
\usetikzlibrary{shapes.geometric}
\RequirePackage{tikz-qtree}
\tikzset{aligned/.style={baseline=(current bounding box.center)}}
\newcommand{\ctree}[1]{%
    \begin{tikzpicture}[sibling distance=1em, level distance=3em, 
        every node/.style = {align=center},
        every internal node/.style = {shape=circle, draw}]
        #1
    \end{tikzpicture}%
}
\newcommand{\tree}[1]{%
    \begin{tikzpicture}[sibling distance=1em, level distance=2em, 
        every node/.style = {align=center}, aligned, remember picture]
        #1
    \end{tikzpicture}%
}
\newcommand{\allctree}[1]{%
    \begin{tikzpicture}[sibling distance=1em, level distance=2em, 
        every node/.style = {align=center, shape=circle, draw}]%
        #1
    \end{tikzpicture}%
}

\newcommand{\ltree}[1]{%
    \begin{tikzpicture}[sibling distance=0.1em, level distance=2.5em, 
        every node/.style = {align=center}, aligned, remember picture]%
        #1%
    \end{tikzpicture}%
}

\newcommand{\lltree}[1]{%
    \begin{tikzpicture}[sibling distance=0.1em, level distance=3.5em, 
        every node/.style = {align=center}, aligned, remember picture]%
        #1%
    \end{tikzpicture}%
}

\newcommand{\ttree}[1]{%
    \begin{tikzpicture}[sibling distance=1em, level distance=3em, 
        every node/.style = {align=center},
        every leaf node/.style = {draw, regular polygon, regular polygon sides=3, anchor=center, inner sep=1pt, rounded corners}]
        #1
    \end{tikzpicture}%
}

\newcommand{\leaf}[1]{%
    \begin{tikzpicture}[every node/.style = {align=center}]
        \Tree [.\ensuremath{#1} ]
    \end{tikzpicture}
}

\makeatletter
\DeclareRobustCommand{\rvdots}{%
  \vbox{
    \baselineskip4\p@\lineskiplimit\z@
    \kern-\p@
    \hbox{.}\hbox{.}\hbox{.}
  }}
\makeatother
 
\newcommand{\trafo}[3][2pt]{%
    \setlength{\tabcolsep}{#1}
    \begin{tabular}{c c c}
        \adjustbox{valign=m}{#2} & 
        \adjustbox{valign=m}{\Large$ \rightsquigarrow $} &
        \adjustbox{valign=m}{#3}
    \end{tabular}%
    
}

\newcommand{\ttrafo}[3][]{%
    \ifthenelse{\isempty{#1}}%
    {\ensuremath{#2 \rightsquigarrow #3}}%
    {\ensuremath{#2 \rightsquigarrow_{#1} #3}}%
}

\newcommand{\tsteptrafo}[4][]{%
    \ifthenelse{\isempty{#1}}%
    {\ensuremath{#3 \rightsquigarrow^{#2} #4}}%
    {\ensuremath{#3 \rightsquigarrow_{#1}^{#2} #4}}%
}
\setcopyright{acmlicensed}
\copyrightyear{2018}
\acmYear{2018}
\acmDOI{XXXXXXX.XXXXXXX}
\acmConference[Conference acronym 'XX]{Make sure to enter the correct
  conference title from your rights confirmation email}{June 03--05,
  2018}{Woodstock, NY}
\acmISBN{978-1-4503-XXXX-X/2018/06}

\begin{document}

\title{Finding Common Mistakes In Modelling With Mathematical Formalisms Using LLMs}

\newcommand{\anonymize}[2]{#1}
\newcommand{\anonAuth}[1]{\anonymize{#1}{Anonymous Author}}
\newcommand{\anonAff}{\anonymize{%
    \affiliation{%
      \institution{Ruhr University Bochum}
      \city{Bochum}
      \country{Germany}
    }
}{%
    \affiliation{%
      \institution{Anonymous Institution}
      \city{City}
      \country{Country}
    }
}}
\newcommand{\anonEmail}[1]{\anonymize{#1}{Anonymous Email}}

\newcommand{\affiliationRUB}{\anonAff}

\author{\anonAuth{Lilian Killich}}
\email{\anonEmail{lilian.killich@rub.de}}

\affiliationRUB

\author{\anonAuth{Marko Schmellenkamp}}
\email{\anonEmail{marko.schmellenkamp@rub.de}}

\affiliationRUB

\author{\anonAuth{Fabian Vehlken}}
\email{\anonEmail{fabian.vehlken@rub.de}}

\affiliationRUB

\author{\anonAuth{Thomas Zeume}}
\email{\anonEmail{thomas.zeume@rub.de}}
\affiliationRUB

\begin{abstract}

  Modelling with mathematical formalisms like logical formulas, mathematical equations, or regular expressions is an important yet challenging task for students of computer science and other STEM disciplines. Identifying common mistakes occurring in this context is an important step towards helping struggling students by providing targeted high-quality feedback, e.g.\ in interactive learning systems.

  We present a tool-supported workflow that allows to (1) identify candidates for common mistakes that explain many student mistakes in large educational data sets, (2) cluster candidates according to similarities, and (3) visualize resulting clusters for instructors and CS education researchers. The visualization is designed to help researchers to identify common modelling mistakes. The candidates for common mistakes are represented by bug fixing transformations that translate incorrect formalizations into correct formalizations; they are generated by an LLM and validated algorithmically.

  We show that this approach works well by reproducing common mistakes in propositional logic modelling that were identified by hand in the literature; showing that, unlike other algorithmic approaches, the LLM-based approach is suitable for very large sets of data; and applying it to multiple other formalisms to showcase it generalizes beyond propositional logic.
\end{abstract}



\maketitle
\pagestyle{plain}

\newcommand{\onlineAppendixRef}{(anonymously published) supplementary material \cite{supplementaryMaterial}\xspace}
\newcommand{\onlineAppendixRefShort}{supplementary material \cite{supplementaryMaterial}\xspace}

\section{Introduction and Motivation}
Learning to model scenarios with mathematical formalisms such as logical formulas, mathematical equations and terms, regular expressions, and many others, is an important aspect when introducing students to mathematical foundations in computer science and other STEM disciplines. A typical learning objective is that students can model scenarios given in natural language within the intended formalism. In this paper we address the question:

\begin{list}{}{\leftmargin=1.5em \rightmargin=1.5em}
\item \textit{Suppose we are given a large data set of student modelling attempts. How can common modelling mistakes be identified?}
\end{list}

Identifying common mistakes is a preliminary step for understanding students' struggles, exploring their invalid conceptions, and ultimately providing advice and feedback in modelling assignments. Knowing common mistakes also helps to craft specific advice and feedback in interactive learning systems.%

We aim for a workflow for identifying common modelling mistakes that supports CS education researchers and educators with adequate algorithmic tools, so that mistakes can be identified from large scale data sets. Our starting point for an algorithmic approach is the observation that the syntax of many mathematical formalisms is specified by a context-free grammar and therefore each formalization is associated with a syntax tree. For such formalisms, modelling mistakes are reflected in differences between the syntax trees of correct and incorrect formalizations. A key idea for answering the above question is that common mistakes can now be found by identifying ``bug fixing transformations'' that explain differences between syntax trees \cite{NeiderSSVZ25}.
We provide examples of this idea for two different formalisms.

\begin{example}[Modelling with propositional logic formulas]\label{example:propositional-modelling-intro}
 A typical assignment when learning propositional logic is the translation of natural language statements such as ``Anna goes to the cinema, but Bob only goes if Celine does.'' into formulas. A correct propositional formula is $ A \wedge (B \to C)$, while a frequent mistake by students is to model the statement by $ A \wedge (C \to B)$. The difference is reflected in the syntax trees:
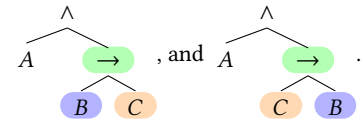
\begin{center}
    \tree{%
        \Tree[.\node (root) {\ensuremath{\land}};
            \node (left) {\ensuremath{A}};
            [.\node (impl) {\ensuremath{\to}}; \node (b) {\ensuremath{B}}; \node (c) {\ensuremath{C}}; ]
        ]
        \coordinate (r_l) at ($(impl.west) + (3pt,0.75pt)$);
        \coordinate (r_r) at ($(impl.east) + (-3pt,0.75pt)$);
        \coordinate (b_l) at ($(b.west) + (3pt,0.75pt)$);
        \coordinate (b_r) at ($(b.east) + (-3pt,0.75pt)$);
        \coordinate (c_l) at ($(c.west) + (3pt,0.75pt)$);
        \coordinate (c_r) at ($(c.east) + (-3pt,0.75pt)$);
        \begin{pgfonlayer}{background}
            \fill[blue,opacity=0.3] \convexpath{b_l,b_r}{5pt};%
            \fill[orange,opacity=0.3] \convexpath{c_l,c_r}{5pt};%
            \fill[green,opacity=0.3] \convexpath{r_l,r_r}{5pt};%
        \end{pgfonlayer}
    }, and 
    \tree{%
        \Tree[.\node (root) {\ensuremath{\land}};
            \node (left) {\ensuremath{A}};
            [.\node (impl) {\ensuremath{\to}}; 
                \node (c) {\ensuremath{C}}; 
                \node (b) {\ensuremath{B}}; 
            ]
        ]
        \coordinate (r_l) at ($(impl.west) + (3pt,0.75pt)$);
        \coordinate (r_r) at ($(impl.east) + (-3pt,0.75pt)$);
        \coordinate (b_l) at ($(b.west) + (3pt,0.75pt)$);
        \coordinate (b_r) at ($(b.east) + (-3pt,0.75pt)$);
        \coordinate (c_l) at ($(c.west) + (3pt,0.75pt)$);
        \coordinate (c_r) at ($(c.east) + (-3pt,0.75pt)$);
        \begin{pgfonlayer}{background}
            \fill[blue,opacity=0.3] \convexpath{b_l,b_r}{5pt};%
            \fill[orange,opacity=0.3] \convexpath{c_l,c_r}{5pt};%
            \fill[green,opacity=0.3] \convexpath{r_l,r_r}{5pt};%
        \end{pgfonlayer}
    }.
\end{center}
 A similar difference occurs for the statement ``The user interface is not working properly only if the database or the back end are not working properly.'' modelled by $ \neg U \to  (\neg D \lor \neg B)$ with frequent mistake $ (\neg D \lor \neg B) \to \neg U$ and the following syntax trees:
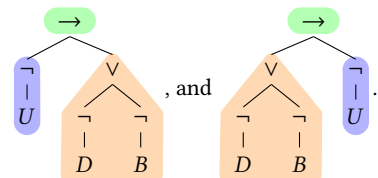
\begin{center}
    \tree{%
        \Tree[.\node (root) {\ensuremath{\to}};
            [.\node (notU) {\ensuremath{\lnot}}; \node (u) {\ensuremath{U}}; ]
            [.\node (or) {\ensuremath{\lor}}; 
                [.\node (notd) {\ensuremath{\lnot}}; \node (d) {\ensuremath{D}}; ]
                [.\node (notb) {\ensuremath{\lnot}}; \node (b) {\ensuremath{B}}; ]
            ]
        ]
        \coordinate (r_l) at ($(root.west) + (3pt,0.75pt)$);
        \coordinate (r_r) at ($(root.east) + (-3pt,0.75pt)$);
        \coordinate (y1_t) at ($(notU.north) + (0,-4pt)$);
        \coordinate (y1_b) at ($(u.south) + (0,4pt)$);
        \coordinate (y2_bl) at ($(d.south west) + (0,2pt)$);
        \coordinate (y2_ml) at ($(notd.west) + (0,2pt)$);
        \coordinate (y2_t) at ($(or.north) + (0,-2pt)$);
        \coordinate (y2_mr) at ($(notb.east) + (0,2pt)$);
        \coordinate (y2_br) at ($(b.south east) + (0,2pt)$);
        \begin{pgfonlayer}{background}
            \fill[green,opacity=0.3] \convexpath{r_l,r_r}{5pt};
            \fill[blue,opacity=0.3] \convexpath{y1_b,y1_t}{5pt};
            \fill[orange,opacity=0.3] \convexpath{y2_bl,y2_ml,y2_t,y2_mr,y2_br}{2.5pt};%
        \end{pgfonlayer}
    }, and 
    \tree{%
        \Tree[.\node (root) {\ensuremath{\to}};
            [.\node (or) {\ensuremath{\lor}}; 
                [.\node (notd) {\ensuremath{\lnot}}; \node (d) {\ensuremath{D}}; ]
                [.\node (notb) {\ensuremath{\lnot}}; \node (b) {\ensuremath{B}}; ]
            ]
            [.\node (notU) {\ensuremath{\lnot}}; \node (u) {\ensuremath{U}}; ]
        ]
        \coordinate (r_l) at ($(root.west) + (3pt,0.75pt)$);
        \coordinate (r_r) at ($(root.east) + (-3pt,0.75pt)$);
        \coordinate (y1_t) at ($(notU.north) + (0,-4pt)$);
        \coordinate (y1_b) at ($(u.south) + (0,4pt)$);
        \coordinate (y2_bl) at ($(d.south west) + (0,2pt)$);
        \coordinate (y2_ml) at ($(notd.west) + (0,2pt)$);
        \coordinate (y2_t) at ($(or.north) + (0,-2pt)$);
        \coordinate (y2_mr) at ($(notb.east) + (0,2pt)$);
        \coordinate (y2_br) at ($(b.south east) + (0,2pt)$);
        \begin{pgfonlayer}{background}
            \fill[green,opacity=0.3] \convexpath{r_l,r_r}{5pt};
            \fill[blue,opacity=0.3] \convexpath{y1_b,y1_t}{5pt};
            \fill[orange,opacity=0.3] \convexpath{y2_bl,y2_ml,y2_t,y2_mr,y2_br}{2.5pt};%
        \end{pgfonlayer}
    }.
\end{center}

 The differences in both examples are captured by a bug fixing transformation that captures the mistake and locally modifies the incorrect trees to correct trees:

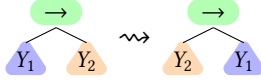
\begin{center}
\trafo{%
    \tree{%
        \Tree[.\node (root) {\ensuremath{\to}}; \node (y1) {\ensuremath{Y_1}}; \node (y2) {\ensuremath{Y_2}}; ]%
        \coordinate (r_l) at ($(root.west) + (3pt,0.75pt)$);
        \coordinate (r_r) at ($(root.east) + (-3pt,0.75pt)$);
        \coordinate (y1_t) at ($(y1.north) + (0,-2pt)$);
        \coordinate (y1_b_l) at ($(y1.south west) + (2pt,4pt)$);
        \coordinate (y1_b_r) at ($(y1.south east) + (-2pt,4pt)$);
        \coordinate (y2_t) at ($(y2.north) + (0,-2pt)$);
        \coordinate (y2_b_l) at ($(y2.south west) + (2pt,4pt)$);
        \coordinate (y2_b_r) at ($(y2.south east) + (-2pt,4pt)$);
        \begin{pgfonlayer}{background}
            \fill[blue,opacity=0.3] \convexpath{y1_b_l,y1_t,y1_b_r}{2.5pt};%
            \fill[orange,opacity=0.3] \convexpath{y2_b_l,y2_t,y2_b_r}{2.5pt};%
            \fill[green,opacity=0.3] \convexpath{r_l,r_r}{5pt};%
        \end{pgfonlayer}
    }%
}{%
    \tree{%
        \Tree[.\node (root) {\ensuremath{\to}}; \node (y2) {\ensuremath{Y_2}}; \node (y1) {\ensuremath{Y_1}}; ]%
        \coordinate (r_l) at ($(root.west) + (3pt,0.75pt)$);
        \coordinate (r_r) at ($(root.east) + (-3pt,0.75pt)$);
        \coordinate (y1_t) at ($(y1.north) + (0,-2pt)$);
        \coordinate (y1_b_l) at ($(y1.south west) + (2pt,4pt)$);
        \coordinate (y1_b_r) at ($(y1.south east) + (-2pt,4pt)$);
        \coordinate (y2_t) at ($(y2.north) + (0,-2pt)$);
        \coordinate (y2_b_l) at ($(y2.south west) + (2pt,4pt)$);
        \coordinate (y2_b_r) at ($(y2.south east) + (-2pt,4pt)$);
        \begin{pgfonlayer}{background}
            \fill[blue,opacity=0.3] \convexpath{y1_b_l,y1_t,y1_b_r}{2.5pt};%
            \fill[orange,opacity=0.3] \convexpath{y2_b_l,y2_t,y2_b_r}{2.5pt};%
            \fill[green,opacity=0.3] \convexpath{r_l,r_r}{5pt};%
        \end{pgfonlayer}
    }%
}
\end{center}

This bug fixing transformation matches the two examples from above as follows: the pattern on the left hand side can be mapped into a subtree of the syntax tree of the incorrect formalizations, and rearranging the subtree according to the right hand side of the transformation yields the correct formalizations (see \cite{NeiderSSVZ25} for precise semantics). This transformation captures the common mistake that students model ``only if'' statements with an implication in the wrong direction. 
It can be used, for instance, to automatically identify this mistake in student attempts and provide feedback \cite{GeckLPSVZ18}.

\end{example}
\begin{example}[Modelling with regular expressions]\label{example:regex-modelling-intro}

    For the assignment ``Provide a \emph{regular expression} for the language of words over the alphabet $ \{a,b,c\} $ in which $ a $ appears as first and last letter, but nowhere else.'' with correct formalization
    $ a (b^*c^*)^* a $ and frequent mistake
    $ a (b^*c^*) a $ with syntax trees

    \begin{center}
    \tree{%
        \Tree[.\node (root) {\ensuremath{\circ}}; 
            \node (leftA) {\ensuremath{a}}; 
            [.\node (innerConc) {\ensuremath{\circ}}; 
                [.\node (leftStar) {\ensuremath{*}}; \node (b) {\ensuremath{b}}; ] 
                [.\node (rightStar) {\ensuremath{*}}; \node (c) {\ensuremath{c}}; ] 
            ] 
            \node (rightA) {\ensuremath{a}}; 
        ]
        \coordinate (conc_bl) at ($(b.south west) + (0,2pt)$);
        \coordinate (conc_ml) at ($(leftStar.west) + (0,2pt)$);
        \coordinate (conc_t) at ($(innerConc.north) + (0,-2pt)$);
        \coordinate (conc_mr) at ($(rightStar.east) + (0,2pt)$);
        \coordinate (conc_br) at ($(c.south east) + (0,2pt)$);
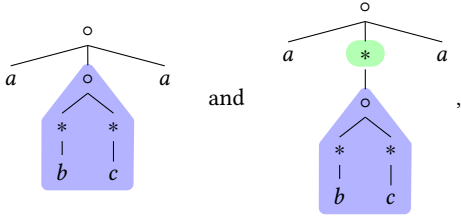
        \begin{pgfonlayer}{background}
            \fill[blue,opacity=0.3] \convexpath{conc_bl,conc_ml,conc_t,conc_mr,conc_br}{2.5pt};%
        \end{pgfonlayer}
    }
    $\quad$and$\quad$
    \tree{%
        \Tree[.\node (root) {\ensuremath{\circ}}; 
            \node (leftA) {\ensuremath{a}}; 
            [.\node (starroot) {\ensuremath{*}}; 
                [.\node (innerConc) {\ensuremath{\circ}}; 
                    [.\node (leftStar) {\ensuremath{*}}; \node (b) {\ensuremath{b}}; ] 
                    [.\node (rightStar) {\ensuremath{*}}; \node (c) {\ensuremath{c}}; ] 
                ] 
            ] 
            \node (rightA) {\ensuremath{a}}; 
        ]
        \coordinate (conc_bl) at ($(b.south west) + (0,2pt)$);
        \coordinate (conc_ml) at ($(leftStar.west) + (0,2pt)$);
        \coordinate (conc_t) at ($(innerConc.north) + (0,-2pt)$);
        \coordinate (conc_mr) at ($(rightStar.east) + (0,2pt)$);
        \coordinate (conc_br) at ($(c.south east) + (0,2pt)$);
        \coordinate (r_l) at ($(starroot.west) + (3pt,0.75pt)$);
        \coordinate (r_r) at ($(starroot.east) + (-3pt,0.75pt)$);
        \begin{pgfonlayer}{background}
            \fill[blue,opacity=0.3] \convexpath{conc_bl,conc_ml,conc_t,conc_mr,conc_br}{2.5pt};%
            \fill[green,opacity=0.3] \convexpath{r_l,r_r}{5pt};%
        \end{pgfonlayer}
    },
    \end{center}
    the difference between the trees can be described by%

    \begin{center}
    \trafo{%
        \tree{%
            \Tree[.\node (root) {\ensuremath{\circ}}; 
                    [.\node (leftStar) {\ensuremath{*}}; \node (y1) {\ensuremath{Y_1}}; ] 
                    [.\node (rightStar) {\ensuremath{*}}; \node (y2) {\ensuremath{Y_2}}; ]             
            ]%
            \coordinate (conc_bl) at ($(y1.south west) + (0,2pt)$);
            \coordinate (conc_ml) at ($(leftStar.west) + (0,2pt)$);
            \coordinate (conc_t) at ($(root.north) + (0,-2pt)$);
            \coordinate (conc_mr) at ($(rightStar.east) + (0,2pt)$);
            \coordinate (conc_br) at ($(y2.south east) + (0,2pt)$);
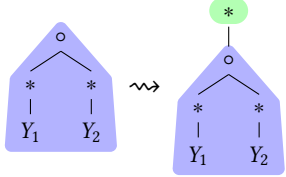
            \begin{pgfonlayer}{background}
                \fill[blue,opacity=0.3] \convexpath{conc_bl,conc_ml,conc_t,conc_mr,conc_br}{2.5pt};%
            \end{pgfonlayer}
        }%
    }{%
        \tree{%
            \Tree[.\node (starroot) {\ensuremath{*}};
                [.\node (root) {\ensuremath{\circ}}; 
                    [.\node (leftStar) {\ensuremath{*}}; \node (y1) {\ensuremath{Y_1}}; ] 
                    [.\node (rightStar) {\ensuremath{*}}; \node (y2) {\ensuremath{Y_2}}; ] 
                ] 
            ]%
            \coordinate (conc_bl) at ($(y1.south west) + (0,2pt)$);
            \coordinate (conc_ml) at ($(leftStar.west) + (0,2pt)$);
            \coordinate (conc_t) at ($(root.north) + (0,-2pt)$);
            \coordinate (conc_mr) at ($(rightStar.east) + (0,2pt)$);
            \coordinate (conc_br) at ($(y2.south east) + (0,2pt)$);
            \coordinate (r_l) at ($(starroot.west) + (3pt,0.75pt)$);
            \coordinate (r_r) at ($(starroot.east) + (-3pt,0.75pt)$);
            \begin{pgfonlayer}{background}
                \fill[blue,opacity=0.3] \convexpath{conc_bl,conc_ml,conc_t,conc_mr,conc_br}{2.5pt};%
                \fill[green,opacity=0.3] \convexpath{r_l,r_r}{5pt};%
            \end{pgfonlayer}
        }%
    }
    \end{center}
\end{example}

\paragraph{Contributions} We present a tool-supported workflow (see Figure \ref{fig:workflow-overview}) that allows to (1) identify candidates for common mistakes by finding  bug fixing transformations that explain many student mistakes in large educational data sets, (2) cluster candidates according to similarities, and (3) visualize resulting clusters for instructors and CS education researchers. The visualization is designed to help researchers to identify common modelling mistakes.  We note that while candidates identified in Step (1) explain frequent structural differences between syntax trees of correct and incorrect formalizations, it is not guaranteed that these differences correspond to actual conceptual mistakes. Therefore, structuring and visualizing the candidates in Steps (2) and (3) is a necessary follow-up step to support researchers and educators to check the candidates and select sensible common mistakes. %

Identifying candidates for common mistakes has so far been a time consuming process for CS education researchers, which required to go over data sets by hand. We provide algorithms that support Steps (1) -- (3) from above, implement them, and thereby reduce the required effort considerably. The main technical challenge is with respect to Step (1). While in \cite{NeiderSSVZ25} a suitable specification language for tree transformations has been crafted and an algorithmic approach to identify transformations from data via a SAT solver was presented, our initial experiments show that this approach does not scale to large educational data sets. Yet, typical such data looks very structured and it is not very hard -- though time consuming -- for humans to find reasonable bug fixing transformations. This motivates our approach to use large language models (LLMs) to find candidate transformations (see Section \ref{section:workflow} for details).

We then employ the resulting tool-supported workflow on educational data sets for logical modelling:
\begin{itemize}
 \item[(a)] On a data set for propositional logic with 6106 pairs of correct and incorrect formalizations, our workflow discovers 
 248 clusters of bug fixing transformations that can help to identify common mistakes. 
 These clusters explain $ 84.44 $\%
 of the pairs in total, whereas common mistakes identified in previous work \cite{SchmellenkampLZ23} explain $ 71.57 $\% of the pairs. Our approach exclusively explains $ 15.03 $\% of the pairs, while $ 2.16 $\% of the pairs are explained exclusively by previous work. When only considering clusters explaining at least $ 0.5 $\% of the pairs, there are 21 clusters which still explain $ 66.34 $\% of pairs; with $ 4.6 $\% explained exclusively.

 \item[(b)] On data sets for modal logic with 12482 pairs of correct and incorrect formalizations, our workflow yields clusters of bug fixing transformations that explain $ 79.39 $\%. When only considering clusters explaining at least $ 0.5 $\% of the pairs, there are 38 clusters for modal logic which explain $ 40.36 $\%. For computation tree logic (CTL) with 7210 pairs, $ 35.89 $\% of the pairs are explained, and $ 19.78 $\% of the pairs are explained by 16 clusters which together explain $0.5$\% of the pairs each.
\end{itemize}

\section{Related Work}
Research on mistakes is conducted for several formalisms. Early work focused on propositional logic \cite{Almstrum99,HermannLKZ12}, with later refinements by Schmellenkamp et al. \cite{SchmellenkampLZ23}, who distinguish linguistic from logical operators to explain discrepancies in earlier findings. Analyses have also been conducted among others for regular expressions \cite{OkuboyejoES2025} and context-free grammars \cite{Pillay2010,SchmellenkampSMZ24}.

A common representation of such mistakes uses bug fixing transformations that describe how incorrect attempts can be systematically corrected. In theoretical domains, this idea has been applied to feedback generation for propositional logic in tutoring systems \cite{GeckLPSVZ18} and extended by \cite{SchmellenkampLZ23}. It has since been generalized to other formalisms, including context-free grammars \cite{SchmellenkampZAKSS25}, demonstrating applicability beyond tree-structured representations. 
Related ideas also appear in work on regular expressions \cite{OkuboyejoES2025}, where feedback is derived from systematic differences between incorrect and correct solutions.

The view that student errors can be explained by systematic rules has a long tradition in intelligent tutoring systems. Brown and Burton \cite{BrownB1978} introduced diagnostic models of procedural bugs in arithmetic, and Sleeman \cite{Sleeman1982} studied how such rules can be inferred for computer-aided instruction. Our notion of mistake candidates as transformations follows this tradition by representing errors as rule-like explanations for families of incorrect formalizations.

Repair-based approaches synthesize transformations from incorrect to correct artifacts, for example in programming education \cite{SinghGS2013,Gulwani2014,AhmedKKKG2018,RiversK2017,XuC2003}. More recent work explores large language models for automated feedback \cite{SilvaC2025,ScholzNSN2025}, showing their potential while also highlighting the need for validation due to possible unreliability or misalignment with instructional goals. We adopt the repair-based perspective while incorporating external algorithmic verification.

\section{A Workflow for Finding Common Mistakes}\label{section:workflow}

\begin{figure*}
    \centering
\noindent
\resizebox{\textwidth}{!}{%
\begin{tikzpicture}[
  >=Latex,
  arrow/.style={->, thick},
  smallarrow/.style={->, thin},
  box/.style={
    draw,
    rounded corners=2pt,
    align=center,
    fill=white
  },
  blob/.style={
    draw,
    ellipse,
    fill=white,
    align=center
  },
  lbl/.style={font=\small},
  batch/.style={
    draw,
    rounded corners=3mm,
    fill=white,
    align=center,
    minimum width=4.25cm,
    minimum height=.78cm,
    lbl
  },
  cluster/.style n args={1}{
    draw,
    ellipse,
    fill=gray!18,
    minimum width=#1,
    minimum height=.55cm,
    inner sep=0mm
  },
  tnode/.style={circle, draw, minimum size=5mm, inner sep=0pt}
]

\node[blob, minimum height=1cm] (dataset) at (0,0)
  {data set};

\node[
  draw,
  fill=gray!6,
  minimum width=18.2cm,
  minimum height=7.1cm,
  anchor=west
] (mainpanel) at (1.5,0) {};

\node[font=\large] at ($(mainpanel.north)+(0,0.42)$)
  {Finding Bug Fixing Transformations};

\draw[arrow] (dataset.east) -- (mainpanel.west);

\node[align=left, anchor=north west] (list)
  at ($(mainpanel.north west)+(0.70,-0.75)$)
{
  $(A\wedge(C\to B),\, A\wedge(B\to C)),$\\[2mm]
  $(\lnot U \to (S \lor T), (S \lor T) \to \lnot U),$\\[2mm]
  $(F\to S,\, S\to F),$\\[2mm]
  \hspace{5mm}$\vdots$\\[2mm]
  $(A\leftrightarrow C,\, A\to C)$
};

\node[above=1mm of list,lbl]{data set};

\draw[
  decorate,
  decoration={brace, amplitude=4pt}
]
  ($(list.north east)+(0,.08)$) --
  ($(list.south east)+(0,-.08)$);
\draw[
  decorate,
  decoration={brace, amplitude=4pt, mirror}
]
  ($(list.north west)+(0,.08)$) --
  ($(list.south west)+(0,-.08)$);

\coordinate (dataBraceTip) at ($(list.east)+(.15,0)$);
\coordinate (dataBraceTipLeft) at ($(list.west)+(.15,0)$);

\node[batch] (b1)
  at ($(mainpanel.north west)+(7.90,-1.05)$)
  {
    $(F\to S,\, S\to F),$\\[0mm]
    $(A\wedge(C\to B),\, A\wedge(B\to C)),$\\[0mm]
    $ \dots $
  };

\node[batch] (b2)
  at ($(mainpanel.north west)+(7.90,-2.20)$)
  {
    $(\lnot U \to (S \lor T), (S \lor T) \to \lnot U),$\\[0mm]
    $\dots$
  };

\node[batch] (b3)
  at ($(mainpanel.north west)+(7.90,-3.55)$)
  {
    $(A\leftrightarrow C,\, A\to C)$\\[0mm]
    $ \dots $
  };

\node at ($(b2)!0.45!(b3)$) {$\vdots$};
\node[lbl] at ($(b1.north)+(0,0.32)$) {randomized batches};

\draw[smallarrow, shorten >= 1mm] (dataBraceTip) -- (b1.west);
\draw[smallarrow, shorten >= 1mm] (dataBraceTip) -- (b2.west);
\draw[smallarrow, shorten >= 1mm] (dataBraceTip) -- (b3.west);

\node[box, minimum width=1.05cm, minimum height=.65cm, lbl] (llm1)
  at ($(mainpanel.north west)+(11.00,-1.05)$)
  {LLM\\Prompt};

\node[box, minimum width=1.05cm, minimum height=.65cm, lbl] (llm2)
  at ($(mainpanel.north west)+(11.00,-2.20)$)
  {LLM\\Prompt};

\node[box, minimum width=1.05cm, minimum height=.65cm, lbl] (llm3)
  at ($(mainpanel.north west)+(11.00,-3.55)$)
  {LLM\\Prompt};

\node at ($(llm2)!0.45!(llm3)$) {$\vdots$};

\draw[smallarrow] (b1.east) -- (llm1.west);
\draw[smallarrow] (b2.east) -- (llm2.west);
\draw[smallarrow] (b3.east) -- (llm3.west);

\node[align=center, anchor=north west, lbl] (transforms)
  at ($(mainpanel.north west)+(12.5,-0.75)$)
{
  \trafo{
      \tree{%
          \Tree[.\ensuremath{\to} \ensuremath{Y_1} \ensuremath{Y_2} ]
      }
  }{
      \tree{%
          \Tree[.\ensuremath{\to} \ensuremath{Y_2} \ensuremath{Y_1} ]
      }
  }, \\
  \trafo{
      \tree{%
          \Tree[.\ensuremath{\leftrightarrow} \ensuremath{Y_1} \ensuremath{Y_2} ]
      }
  }{
      \tree{%
          \Tree[.\ensuremath{\to} \ensuremath{Y_1} \ensuremath{Y_2} ]
      }
  }, \\

  $\vdots$
};
\node[above=1mm of transforms] {
        set of transformations
};

\draw[
  decorate,
  decoration={brace, amplitude=3pt}
]
  ($(transforms.south west)+(0.2,-.08)$) --
  ($(transforms.north west)+(0.2,.08)$);

\coordinate (transBraceTip) at ($(transforms.west)+(.10,0)$);

\draw[
  decorate,
  decoration={brace, amplitude=3pt}
]
  ($(transforms.north east)+(0.1,.08)$) --
  ($(transforms.south east)+(0.1,-.08)$);

\draw[smallarrow, shorten >= 1mm] (llm1.east) -- (transBraceTip);
\draw[smallarrow, shorten >= 1mm] (llm2.east) -- (transBraceTip);
\draw[smallarrow, shorten >= 1mm] (llm3.east) -- (transBraceTip);

\node[blob, minimum width=2.85cm, minimum height=.85cm, lbl] (remain)
  at ($(transforms.south)+(0,-1.5)$)
  {remaining data set with\\unexplained pairs};

\draw[smallarrow]
  ($(transforms.south)+(0,-0.1)$) --
  node[right,lbl,align=left] {apply to\\ input data set}
  (remain.north);

\node[
  draw,
  regular polygon,
  regular polygon sides=6,
  shape aspect=1.5,
  align=center,
  lbl,
  fill=white,
  inner sep=-0.8mm
] (term)
  at ($(mainpanel.south)+(0,1.05)$)
  {termination\\criterion\\satisfied?};

\draw[smallarrow] (remain.west) -| (term.north);

\node[
  box,
  fill=green!10,
  minimum width=1.95cm,
  minimum height=1.05cm,
  lbl
] (out)
  at ($(mainpanel.east)+(-0.55,0)$)
  {\emph{output}: \\set of collected\\transformations};

\draw[smallarrow]
  (term.east) -|
  node[pos=.05, below, lbl] {yes}
  (out.south);

\draw[smallarrow]
  (term.west) -|
  node[pos=.05, below, lbl] {no}
  ($(dataBraceTipLeft)+(-0.6,0)$)
  |-
  ($(dataBraceTipLeft)+(-0.3,0)$)
  ;

  \node[draw,ellipse,fill=white] at ($(term.west)+(-6,0)$) {repeat};

\node[
  draw,
  fill=orange!8,
  minimum width=5.2cm,
  minimum height=7.1cm,
  anchor=west
] (clusterpanel)
  at ($(mainpanel.east)+(1.25,0)$) {};

\node[font=\large] at ($(clusterpanel.north)+(0,0.42)$)
  {Clustering};

\draw[arrow] (out.east) -- (clusterpanel.west);

\node[cluster={1.15cm}] (c1) at ($(clusterpanel.north west)+(1.25,-3.25)$) {
  \begin{tikzpicture}[
  every node/.style={circle, draw, minimum size=1mm, inner sep=0pt},
  level distance=2mm,
  sibling distance=3mm
]

\node (L) {}
  child { node {} }
  child { node {} };

\node (R) at (1,0) {}
  child { node {} }
  child { node {} };

\node[draw=none] at ($(L)!0.5!(R)$) {$\rightsquigarrow$};

\end{tikzpicture}
};

\node[cluster] (c4) at ($(clusterpanel.north west)+(2,-2.25)$) {
  \begin{tikzpicture}[
  every node/.style={circle, draw, minimum size=1mm, inner sep=0pt},
  level distance=2mm,
  sibling distance=3mm
]

\node (L) {};

\node (R) at (1,0) {};

\node[draw=none] at ($(L)!0.5!(R)$) {$\rightsquigarrow$};

\end{tikzpicture}
};

\node[cluster]  (c3) at ($(c4.north west)+(-0.4,1)$) {
  \begin{tikzpicture}[
  every node/.style={circle, draw, minimum size=1mm, inner sep=0pt},
  level distance=2mm,
  sibling distance=3mm
]

\node (L) {}
  child { node {} }
  child { node {} };

\node (R) at (1,0) {};

\node[draw=none] at ($(L)!0.5!(R)$) {$\rightsquigarrow$};

\end{tikzpicture}
};

\node[cluster] (c2) at ($(c4.north east)+(0.4,1)$) {
  \begin{tikzpicture}[
  every node/.style={circle, draw, minimum size=1mm, inner sep=0pt},
  level distance=2mm,
  sibling distance=3mm
]

\node (L) {};

\node (R) at (1,0) {}
  child { node {} }
  child { node {} };

\node[draw=none] at ($(L)!0.5!(R)$) {$\rightsquigarrow$};

\end{tikzpicture}
};

\node[cluster] (c5) at ($(clusterpanel.north west)+(4.00,-2.15)$) {
  \begin{tikzpicture}[
  every node/.style={circle, draw, minimum size=1mm, inner sep=0pt},
  level distance=2mm,
  sibling distance=3mm
]

\node (L) {}
  child { node {} }
  child { node {} };

\node (R) at (1,0) {}
  child { node {} }
  child { node {} };

\node[draw=none] at ($(L)!0.5!(R)$) {$\rightsquigarrow$};

\end{tikzpicture}
};

\draw[smallarrow] (c4) -- (c2);
\draw[smallarrow] (c4) -- (c3);

\node[cluster, align=center, lbl] (bm)
  at ($(clusterpanel.south)+(0,.5)$)
  {
  \begin{tikzpicture}[
  every node/.style={circle, draw, minimum size=1mm, inner sep=0pt},
  level distance=2mm,
  sibling distance=3mm
]

\node (L) {}
  child { node {} }
  child { node {} };

\node (R) at (1,0) {}
  child { node {} }
  child { node {} };

\node[draw=none] at ($(L)!0.5!(R)$) {$\rightsquigarrow$};

\end{tikzpicture}
  };

\node[cluster, align=center, lbl] (bmlc)
  at ($(bm.north west)+(-0.5,1)$)
  {
  \begin{tikzpicture}[
  every node/.style={circle, draw, minimum size=1mm, inner sep=0pt},
  level distance=2mm,
  sibling distance=3mm
]

\node (L) {}
  child { node {} }
  child { node {}
    child { node {} }
    child { node {} }
  };

\node (R) at (1,0) {}
  child { node {} }
  child { node {} };

\node[draw=none] at ($(L)!0.5!(R)$) {$\rightsquigarrow$};

\end{tikzpicture}
  };

\node[cluster, align=center, lbl] (bmrc)
  at ($(bm.north east)+(0.5,1)$)
  {
  \begin{tikzpicture}[
  every node/.style={circle, draw, minimum size=1mm, inner sep=0pt},
  level distance=2mm,
  sibling distance=3mm
]

\node (L) {}
  child { node {} }
  child { node {}
    child { node {} }
    child { node {} }
  };

\node (R) at (1,0) {}
  child { node {} }
  child { node {} };

\node[draw=none] at ($(L)!0.5!(R)$) {$\rightsquigarrow$};

\end{tikzpicture}
  };

\node[cluster, align=center, lbl] (bmrcc)
  at ($(bmrc.north)+(0,1)$)
  {
  \begin{tikzpicture}[
  every node/.style={circle, draw, minimum size=1mm, inner sep=0pt},
  level distance=2mm,
  sibling distance=3mm
]

\node (L) {}
  child { node {} }
  child { node {}
    child { node {} }
    child { node {} }
  };

\node (R) at (1,0) {}
  child { node {} }
  child { node {}
    child { node {} }
    child { node {} }
  };

\node[draw=none] at ($(L)!0.5!(R)$) {$\rightsquigarrow$};

\end{tikzpicture}
  };

\draw[smallarrow] (bmrc) -- (bmrcc);
\draw[smallarrow, bend right=5] (bmrc) to (bmlc);
\draw[smallarrow, bend right=5] (bmlc) to (bmrc);
\draw[smallarrow] (bm) -- (bmrc);
\draw[smallarrow] (bm) -- (bmlc);

\node[
  draw,
  fill=blue!9,
  minimum width=4.4cm,
  minimum height=7.1cm,
  anchor=west
] (vispanel)
  at ($(clusterpanel.east)+(1,0)$) {};

\node[font=\large] at ($(vispanel.north)+(0,0.42)$)
  {Visualization};

\draw[arrow] (clusterpanel.east) -- (vispanel.west);

\node[font=\large] (hiertable) at ($(vispanel.north)+(0,-3.75)$)
  {
    \begin{tikzpicture}[scale=.9]

  Vertical divider
  \draw (1.5,7) -- (1.5,0);
  \draw (0,6.5) -- (4,6.5);

  \draw[decorate, decoration={snake, amplitude=0.5mm, segment length=4mm}]
    (0,6) -- (1.4,6);
  \draw[decorate, decoration={snake, amplitude=0.5mm, segment length=4mm}]
    (1.6,6) -- (4,6);

  \draw[decorate, decoration={snake, amplitude=0.5mm, segment length=4mm}]
    (0.5,5.5) -- (1.4,5.5);
  \draw[decorate, decoration={snake, amplitude=0.5mm, segment length=4mm}]
    (1.6,5.5) -- (4,5.5);

  \draw[decorate, decoration={snake, amplitude=0.5mm, segment length=4mm}]
    (0,4.8) -- (1.4,4.8);
  \draw[decorate, decoration={snake, amplitude=0.5mm, segment length=4mm}]
    (1.6,4.8) -- (4,4.8);

  \draw[decorate, decoration={snake, amplitude=0.5mm, segment length=4mm}]
    (0.5,4.3) -- (1.4,4.3);
  \draw[decorate, decoration={snake, amplitude=0.5mm, segment length=4mm}]
    (1.6,4.3) -- (4,4.3);

  \draw[decorate, decoration={snake, amplitude=0.5mm, segment length=4mm}]
    (0.5,3.8) -- (1.4,3.8);
  \draw[decorate, decoration={snake, amplitude=0.5mm, segment length=4mm}]
    (1.6,3.8) -- (4,3.8);

  \draw[decorate, decoration={snake, amplitude=0.5mm, segment length=4mm}]
    (1,3.3) -- (1.4,3.3);
  \draw[decorate, decoration={snake, amplitude=0.5mm, segment length=4mm}]
    (1.6,3.3) -- (4,3.3);

  \draw[decorate, decoration={snake, amplitude=0.5mm, segment length=4mm}]
    (0,2.6) -- (1.4,2.6);
  \draw[decorate, decoration={snake, amplitude=0.5mm, segment length=4mm}]
    (1.6,2.6) -- (4,2.6);

  \draw[decorate, decoration={snake, amplitude=0.5mm, segment length=4mm}]
    (0,1.9) -- (1.4,1.9);
  \draw[decorate, decoration={snake, amplitude=0.5mm, segment length=4mm}]
    (1.6,1.9) -- (4,1.9);

  \draw[decorate, decoration={snake, amplitude=0.5mm, segment length=4mm}]
    (0,1.2) -- (1.4,1.2);
  \draw[decorate, decoration={snake, amplitude=0.5mm, segment length=4mm}]
    (1.6,1.2) -- (4,1.2);

  \draw[decorate, decoration={snake, amplitude=0.5mm, segment length=4mm}]
    (0.5,0.7) -- (1.4,0.7);
  \draw[decorate, decoration={snake, amplitude=0.5mm, segment length=4mm}]
    (1.6,0.7) -- (4,0.7);

  \draw[decorate, decoration={snake, amplitude=0.5mm, segment length=4mm}]
    (0.5,0.2) -- (1.4,0.2);
  \draw[decorate, decoration={snake, amplitude=0.5mm, segment length=4mm}]
    (1.6,0.2) -- (4,0.2);
\end{tikzpicture}
  };
\node[lbl] at ($(hiertable.north)+(0,0.2)$) {Hierarchical clusters table:};

\end{tikzpicture}
}
     \caption{Illustration of the workflow for finding bug fixing transformations, clustering them, and visualizing the clusters.}\label{fig:workflow-overview}
\end{figure*}
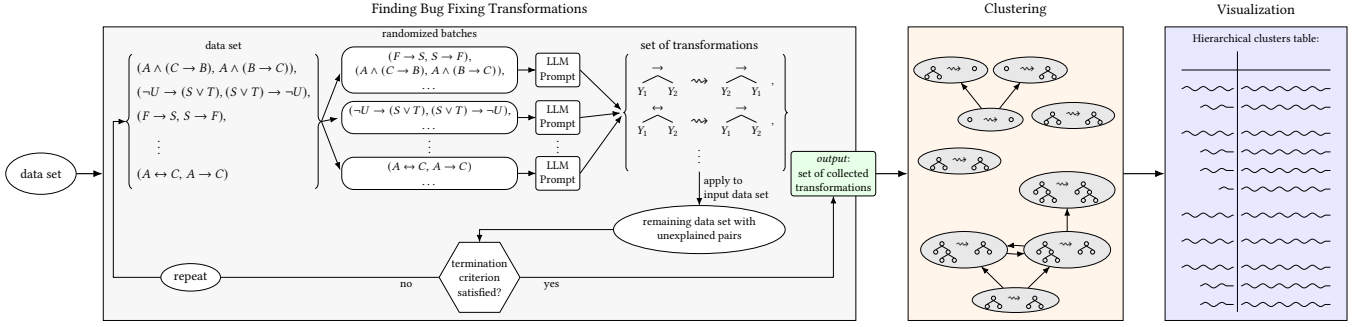

In this section we describe our tool-supported workflow (see Figure~\ref{fig:workflow-overview}) that allows to (1) identify candidates for common mistakes by finding bug fixing transformations that explain many student mistakes in large educational data sets, (2) cluster candidates according to similarities, and (3) visualize resulting clusters for instructors and CS education researchers. Step (1) uses an iterative process to generate candidates via an LLM and validates them algorithmically (see Section \ref{section:bug-fixing-transformations}). For Step (2), we compute a correlation graph between candidate transformations (see Section \ref{section:clustering}).

The workflow is applicable to various data sets. It can be used, for example, on educational data sets for modelling with logical formulas (propositional formulas, modal formulas, temporal formulas, \dots), regular expressions, mathematical expressions etc. Another use case is the application to multiple data sets for the same formalism to discover more specific mistakes. As an example, a partitioned data set for propositional modelling into smaller data sets according to \emph{linguistic operators} occurring in natural language statements has been presented by  Schmellenkamp et al. \cite{SchmellenkampLZ23}. Applying our workflow to each of these smaller data sets results in candidates for common mistakes for each linguistic operator.

\subsection{Finding Bug Fixing Transformations}\label{section:bug-fixing-transformations}
The goal of the first step of our workflow is, given a data set containing pairs of  incorrect and correct formalizations, to find transformations that explain (frequent) structural differences between the correct and incorrect formalizations. More formally,
the input is a data set  $\calD$ which contains pairs $(t, t^*)$ of syntax trees $t$ and $t^*$ representing incorrect and correct formalizations. The goal is to identifiy a small set $\calT$ of bug fixing transformations that explain the difference between as many pairs in $\calD$ as possible. Here, roughly speaking, a  \emph{bug fixing transformation} is of the form  $\sigma \rightsquigarrow \sigma^*$ for two tree patterns $\sigma$ and $\sigma^*$. Such a transformation \emph{explains} a pair $(t, t^*)$ if the pattern $\sigma$ can be found in $t$ and rearranging $t$ according to the shape of $\sigma^*$ yields the tree $t^*$ (see Examples \ref{example:propositional-modelling-intro} and \ref{example:regex-modelling-intro} as well as \cite{NeiderSSVZ25} for details). Unfortunately,  finding a small set of bug fixing transformations that explains a large data set is  \NP-hard \cite{NeiderSSVZ25} and SAT-solving based approaches do not scale well.

Our algorithmic approach (see Algorithm \ref {algorithm:finding-transformations} for pseudo code) is to generate candidates for bug fixing transformations with an LLM,  validate those candidates algorithmically, and collect transformations that explain a $\delta$-fraction of the initial data set in the output set $\calT$. This procedure is iterated until no new transformation has been found for $k$ iterations. In our experiments we used $\delta = 0.01$ and $k = 10$, see Section \ref{section:evaluation} for details.

\makeatletter
\newcommand{\INPUT}{\item[\algorithmicinput]}
\newcommand{\algorithmicinput}{\textbf{Input:}}
\newcommand{\OUTPUT}{\item[\algorithmicoutput]}
\newcommand{\algorithmicoutput}{\textbf{Output:}}
\makeatother
\begin{algorithm}
    \caption{Finding bug fixing transformations}\label{algorithm:finding-transformations}
    \begin{algorithmic}[1]
        \INPUT A set $\calD$ of pairs $(t, t^*)$ of syntax trees
        \OUTPUT A set $\calT$ of bugfixing transformations

        \STATE $ \widehat{\calD} := \calD$
        \STATE $ \calT := \emptyset $

        \WHILE {termination criterion is not fulfilled}
            \STATE $ \calB_1, \dots, \calB_b := $ randomly split $ \widehat{\calD}$ into batches of size $s$
            \FORALL{batches $ \calB_i $}
                \STATE prompt LLM to find transformations
                \STATE add those transformations to $ \calT $  which explain more than a $\delta$-fraction of $\calD$
            \ENDFOR
            \STATE $ \widehat{\calD} := $ pairs in $ \widehat{\calD} $ not explained by any transformation in $ \calT $
        \ENDWHILE
        \RETURN $ \calT $
    \end{algorithmic}

\end{algorithm}%

We provide some details. In each iteration, the data set at hand is split into equally sized randomized batches of size $s$ (in our experiments, we use $s \sim 50$), to accommodate limits of used LLMs. Each of the batches is put into a prompt that also contains a brief description of the transformation language, a few sample specifications for special cases, and a description of the task to find transformations. The prompt that was used can be found in the \onlineAppendixRef. Via this prompt, the LLM provides a list of transformations for each one of the batches. The number of desired transformations for each batch is configurable (in our experiments, we use $ \lfloor 0.2\cdot s \rfloor $). For each transformation found by the LLM, the algorithm tests which fraction of pairs $(t, t^*)$ it can explains, i.e.\ pairs for which $t$ can be transformed into $t^*$ by applying the transformation. If a transformation explains more than a $\delta$-fraction of the data, it is added to the output $\calT$.

\subsection{Clustering Bug Fixing Transformations}\label{section:clustering}
The goal of the second step of the workflow is to structure the list of bug fixing transformations resulting from the first step for CS education researchers and educators. The idea is to cluster transformations according to ``similarity'' and to sort those clusters by the number of explained data points.
The idea is that transformations in the list may explain similar kinds of mistakes or one transformation may explain all mistakes also explained by another transformation. \fvm{I don't understand the point of this}
As an example, consider the transformations
\begin{center}
\midlabelled{$ \rho_1 $}{%
\ttrafo{
    \tree{%
        \Tree[.\ensuremath{\leftrightarrow} \ensuremath{Y_1} \ensuremath{Y_2} ]
    }
}{
    \tree{%
        \Tree[.\ensuremath{\lnot} [.\ensuremath{\leftrightarrow} \ensuremath{Y_1} \ensuremath{Y_2} ] ]
    }
}}
and 
\midlabelled{$ \rho_2 $}{%
\ttrafo{
    \tree{%
        \Tree[.\ensuremath{Y_1} ]
    }
}{
    \tree{%
        \Tree[.\ensuremath{\lnot} \ensuremath{Y_1} ]
    }
}}.

\end{center}

\noindent Both explain mistakes where a negation $ \lnot $ has been forgotten, but the first one is more specific: it only explains pairs in which the negation is missing in front of a bi-implication $ \leftrightarrow $. The first may explain mistakes that are specific when modelling natural language statements that involve negations and equivalences, while the second can explain natural language statements with negations more broadly.

Our algorithmic approach for finding similar transformations is to compute, for each pair $\rho_1$ and $\rho_2$ of transformations, the fraction of data points explained by $\rho_1$ that are also explained by $\rho_2$. More precisely, denote by $ E_\rho(\calD) $ the set of pairs in $ \calD $ that are explained by $ \rho $. We compute a \emph{correlation graph} $ \calG_\calD $ for the transformations found for data set $\calD$ that has a node for each transformation found in Step (1). For each pair $\rho_1$ and $\rho_2$ of transformations, the graph has an edge $(\rho_1, \rho_2)$ of weight $\dfrac{|E_{\rho_1}(\calD) \cap E_{\rho_2}(\calD)|}{|E_{\rho_2}(\calD)|}$. Thus, the weight of $(\rho_1, \rho_2)$ is the fraction of the data points explained by $\rho_2$ which are also explained by $\rho_1$. In particular, a value of $ 1 $ means $ \rho_1 $ explains all pairs of $ \calD $ explained by $ \rho_2 $  (and possibly more) and a value of $0$ means that it explains none. A \emph{cluster} within a correlation graph is a set of nodes that are connected by a path consisting of edges of non-zero weight.

Consider the following illustrative toy example of a correlation graph for a small hand-chosen data set.

\begin{example}\label{example:correlation-graph}
    Consider the following five transformations

    \scalebox{.7}{%
    \midlabelled{$ \rho_1 $}{
        \trafo[0mm]{
            \tree{%
                \Tree[.\ensuremath{\to}
                    \ensuremath{Y_1}
                    \ensuremath{Y_2}
                ]
            }
        }{
            \tree{%
                \Tree[.\ensuremath{\to}
                    \ensuremath{Y_2}
                    \ensuremath{Y_1}
                ]
            }
        }
    }},
    \scalebox{.7}{%
    \midlabelled{$ \rho_2 $}{
        \trafo[0mm]{
            \tree{%
                \Tree[.\ensuremath{\land}
                    \ensuremath{Y_1}
                    [.\ensuremath{\to}
                        \ensuremath{Y_2}
                        \ensuremath{Y_3}
                    ]
                ]
            }
        }{
            \tree{%
                \Tree[.\ensuremath{\land}
                    \ensuremath{Y_1}
                    [.\ensuremath{\to}
                        \ensuremath{Y_3}
                        \ensuremath{Y_2}
                    ]
                ]
            }
        }
    }},

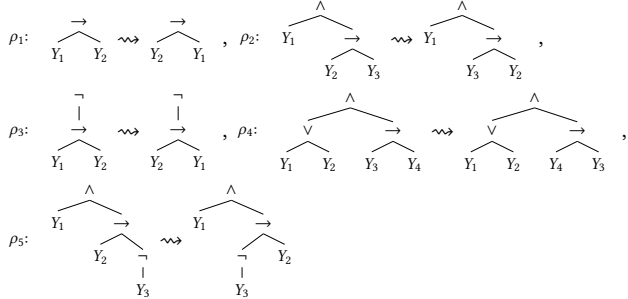
    \scalebox{.7}{%
    \midlabelled{$ \rho_3 $}{
        \trafo[0mm]{
            \tree{%
                \Tree[.\ensuremath{\lnot}
                    [.\ensuremath{\to}
                        \ensuremath{Y_1}
                        \ensuremath{Y_2}
                    ]
                ]
            }
        }{
            \tree{%
                \Tree[.\ensuremath{\lnot}
                    [.\ensuremath{\to}
                        \ensuremath{Y_2}
                        \ensuremath{Y_1}
                    ]
                ]
            }
        }
    }},
    \scalebox{.7}{%
    \midlabelled{$ \rho_4 $}{
        \trafo[0mm]{
            \tree{%
                \Tree[.\ensuremath{\land}
                    [.\ensuremath{\lor}
                        \ensuremath{Y_1}
                        \ensuremath{Y_2}
                    ]
                    [.\ensuremath{\to}
                        \ensuremath{Y_3}
                        \ensuremath{Y_4}
                    ]
                ]
            }
        }{
            \tree{%
                \Tree[.\ensuremath{\land}
                    [.\ensuremath{\lor}
                        \ensuremath{Y_1}
                        \ensuremath{Y_2}
                    ]
                    [.\ensuremath{\to}
                        \ensuremath{Y_4}
                        \ensuremath{Y_3}
                    ]
                ]
            }
        }
    }},

    \scalebox{.7}{%
    \midlabelled{$ \rho_5 $}{
        \trafo[0mm]{
            \tree{%
                \Tree[.\ensuremath{\land}
                    \ensuremath{Y_1}
                    [.\ensuremath{\to}
                        \ensuremath{Y_2}
                        [.\ensuremath{\lnot}
                            \ensuremath{Y_3}
                        ]
                    ]
                ]
            }
        }{
            \tree{%
                \Tree[.\ensuremath{\land}
                    \ensuremath{Y_1}
                    [.\ensuremath{\to}
                        [.\ensuremath{\lnot}
                            \ensuremath{Y_3}
                        ]
                        \ensuremath{Y_2}
                    ]
                ]
            }
        }
    }}

    \noindent
    and data set $ \calS $ consisting of the following (incorrect, correct)-pairs:%
        \begin{align*}
            t_1 &= (A \to B, B \to A),\\ 
            t_2 &= (\lnot A \to B, B \to \lnot A),\\ %
            t_3 &= ((A \lor \lnot D) \land (B \to C), (A \lor \lnot D) \land (C \to B)),\\
            t_4 &= (\lnot((A \lor B) \to (C \to D)), \lnot((C \to D) \to (A \lor B))),\\ %
            t_5 &= ((A \lor B) \land (C \to D), (A \lor B) \land (D \to C)),\\ %
            t_6 &= (A \land (B \to \lnot C), A \land (\lnot C \to B)) %
        \end{align*}
    We first compute the sets $ E_{\rho_i}(\calS) $ of pairs explained by each transformation $ \rho_i $:
    \begin{align*}
        E_{\rho_1}(\calS) &= \{t_1, t_2, t_3, t_4, t_5, t_6\},\\
        E_{\rho_2}(\calS) &= \{t_3, t_5, t_6\},\\
        E_{\rho_3}(\calS) &= \{t_4\},\\
        E_{\rho_4}(\calS) &= \{t_3, t_5\},\\
        E_{\rho_5}(\calS) &= \{t_6\}.
    \end{align*}

    The resulting correlation graph is depicted below. The edges are labelled with their weight, i.e.\ the fraction of pairs explained by the target transformation that are also explained by the source transformation. Self-loops, edges with weight $ 0 $, and ``transitive edges'' $ (\rho_1, \rho_4), (\rho_4, \rho_1), (\rho_1, \rho_5), (\rho_5, \rho_1) $ are omitted for clarity.

    \begin{center}
        \small
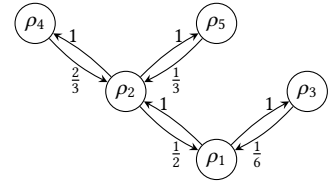
        \begin{tikzpicture}[
    scale=0.6,
    node distance=1.8cm and 2.2cm,
    every node/.style={circle, draw, inner sep=2pt},
    edge/.style={->, >=stealth}
]

\node (r1) at (0,0) {$\rho_1$};

\node (r2) at (-2,1.5) {$\rho_2$};
\node (r3) at ( 2,1.5) {$\rho_3$};

\node (r4) at (-4,3) {$\rho_4$};
\node (r5) at (0,3) {$\rho_5$};

\draw[edge, bend right=8]  (r1) to node[draw=none, circle=none, pos=0.95, right] {$1$} (r2);
\draw[edge, bend right=10] (r2) to node[draw=none, circle=none, pos=0.9, left, xshift=1mm, yshift=-1mm] {$\frac{1}{2}$} (r1);

\draw[edge, bend left=8] (r1) to node[draw=none, circle=none, pos=0.95, left] {$1$} (r3);
\draw[edge, bend left=10] (r3) to node[draw=none, circle=none, pos=0.9, right, xshift=-1mm, yshift=-1mm] {$\frac{1}{6}$} (r1);

\draw[edge, bend right=8]  (r2) to node[draw=none, circle=none, pos=0.95, right] {$1$} (r4);
\draw[edge, bend right=10] (r4) to node[draw=none, circle=none, near end, left, xshift=1mm, yshift=-1mm] {$\frac{2}{3}$} (r2);

\draw[edge, bend left=8] (r2) to node[draw=none, circle=none, pos=0.95, left] {$1$} (r5);
\draw[edge, bend left=10] (r5) to node[draw=none, circle=none, near end, right, xshift=-1mm, yshift=-1mm] {$\frac{1}{3}$} (r2);

\end{tikzpicture}
    \end{center}
    As the correlation graph indicates, $ \rho_1, \dots, \rho_5 $ form a cluster. \qed
\end{example}

We use correlation graphs to provide visualizations of how bug fixing transformations found in Step (1) relate to each other. In our experiments, all correlation graphs can be partitioned into clusters of a very simple shape: (a) \emph{single-transformation clusters} consisting of a single transformation that only explains data points that no other transformation explains; (b) \emph{clusters with equivalent transformations} consisting of transformations that all explain the same data points; and (c) \emph{hierarchical clusters} whose transformations can be arranged in a tree, such that if a transformation explains data points $\calD' \subseteq \calD$ then each of its children explains only data points $\calD'' \subseteq \calD'$. In particular, clusters of type (b) and (c) are helpful for CS education researchers to identify transformations representative for common mistakes and to distinguish ``high-level'' from ``low-level'' mistakes.

\section{Evaluation}\label{section:evaluation}
In this section we employ our tool-supported workflow exemplarily on educational data sets for logical modelling. In all experiments we use the workflow to find transformations, evaluate how well they cover data, and present a selection of transformations that explain large fractions of the data.
Detailed tables on all experiments can be found in the \onlineAppendixRefShort.

In a first set of experiments we use our workflow on a data set for modelling in propositional logic that has been used by Schmellenkamp et al. \cite{SchmellenkampLZ23} to identify common mistakes by hand. Here we apply our workflow to a partition of the data set according to linguistic operators used in natural language statements as well as to the full data set. The objective is to see whether our workflow works as well as the by-hand-approach used in previous work.

Then, in a second set of experiments, we apply our workflow to data sets for modelling with modal logic and computation tree logic (CTL). Here the objective is to see, exemplarily, whether the workflow scales to other formalisms.

We first describe the experimental setup and then the results for the experiments sketched above.

For finding bug fixing transformations in Step (1) of our workflow, we use the GPT-OSS-120B model \cite{openai2025gptoss120bgptoss20bmodel}. Initial experiments showed that this model is powerful enough, but small enough to be locally deployable. As parameters we choose a batch size of 50, due to the token limitations of the model.
In all experiments, the prompt asks for transformations that explain at least 20\% of the pairs in a batch. For the propositional data set grouped by linguistic operators, found transformations additionally need to explain  $\delta = 1 $\% of all data points. The iterative process is terminated if no new transformation is chosen for 10 iterations.

\subsection{Results for propositional logic}%

The propositional data set from \cite{SchmellenkampLZ23} contains $ 6106 $ pairs with $ 1572 $ of them being distinct. This data set comes with a partition into $34$ subdata sets for different linguistic operators. The median and average sizes of the subdata sets are $ 65.5 $ and $ 177.8 $, respectively.

\subsubsection*{Results for linguistic operator data sets}
Across the subdata sets, the found transformations explain at least as many
pairs as the hand-picked mistakes in a substantial fraction of cases.

For $>64$\% of the lingustic operator data sets, our workflow finds bug fixing transformations transformations that explain more data points than the mistakes identified in \cite{SchmellenkampLZ23}. 
For $>14\%$ of the subdata sets, the transformations explain exactly as many pairs as the hand-picked mistakes.

\subsubsection*{Results for the full propositional data set}
\begin{table*}
\caption{Exemplary excerpt of the table of hierarchical bug-fixing transformation clusters that cover most of the pairs explained by the common mistake that a disjunction was erroneously used instead of a conjunction (as identified in \cite{SchmellenkampLZ23}). For each transformation, a simple representative example is provided. For the mistake as a whole, each transformation, and the entire cluster, the number \# and distinct number \#dst of pairs are shown. The common mistake occurrs in $5.47$\% of all pairs in the full data set ($ 3.88 $\% distinct). Of these, the complete bug fixing transformation cluster found by our workflow explains $ 330 $ ($ 98.8 $\%). The fraction of the $ 334 $ pairs each transformation explains is also shown (columns \% and \%dst, respectively). A table with all bug fixing transformation clusters found by our workflow is presented in the~\mbox{\onlineAppendixRef}.}\label{tab:PropositionalUngroupedSorted-errors-trafos_main}
    \begin{tabular}{lrrrr | l l}
        \toprule
        Hand-picked mistake / transformations & \# & \% & \#dst & \%dst & \multicolumn{2}{l}{example of (incorrect, correct) formula pairs} \\
        \midrule
        \textbf{Mistake: disjunction instead of conjunction} & \textbf{334} & \textbf{5.47\%} & \textbf{61} & \textbf{3.88\%} \\
        \textbf{\quad Transformations combined} & \textbf{330} & \textbf{98.80\%} & \textbf{58} & \textbf{95.08\%} \\
        \quad $\lnot \left(Y_{1} \land \ldots \land Y_{n}\right)$ $\rightsquigarrow$ $\lnot Y_{1} \land \ldots \land \lnot Y_{n}$ & 201 & 60.18\% & 23 & 37.70\% 
        & $ \lnot(A \land B \land C),$&$ \lnot A \land \lnot B \land \lnot C $
        \\
        \quad \quad $\lnot \left(Y_{1} \land Y_{2}\right)$ $\rightsquigarrow$ $\lnot Y_{1} \land \lnot Y_{2}$ & 153 & 45.81\% & 16 & 26.23\% 
        & $ \lnot(A \land B),$&$ \lnot A \land \lnot B $
        \\
        \quad \quad \quad $\lnot \left(Y_{1} \land Y_{2}\right) \rightarrow Y_{3}$ $\rightsquigarrow$ $\left(\lnot Y_{1} \land \lnot Y_{2}\right) \rightarrow Y_{3}$ & 150 & 44.91\% & 15 & 24.59\% 
        & $ \lnot(A \land B) \to C,$&$ (\lnot A \land \lnot B) \to C $
        \\
        \quad $Y_{1} \lor \ldots \lor Y_{n}$ $\rightsquigarrow$ $Y_{1} \land \ldots \land Y_{n}$ & 128 & 38.32\% & 34 & 55.74\% 
        & $ A \lor B \lor C,$&$ A \land B \land C $
        \\
        \quad \quad $Y_{1} \lor Y_{2}$ $\rightsquigarrow$ $Y_{1} \land Y_{2}$ & 118 & 35.33\% & 29 & 47.54\% 
        & $ A \lor B,$&$ A \land B $
        \\
        \quad \quad \quad $\lceil$ $=$ $\left(\lnot Y_{1} \lor \lnot Y_{2}\right) \rightarrow Y_{3}$ $\rightsquigarrow$ $\left(\lnot Y_{1} \land \lnot Y_{2}\right) \rightarrow Y_{3}$ & 35 & 10.48\% & 9 & 14.75\% 
        & $ (\lnot A \lor \lnot B) \to C,$&$ (\lnot A \land \lnot B) \to C $
        \\
        \quad \quad \quad $\lfloor$ $=$ $\lnot Y_{1} \lor \lnot Y_{2}$ $\rightsquigarrow$ $\lnot Y_{1} \land \lnot Y_{2}$ & 35 & 10.48\% & 9 & 14.75\% 
        & $ (\lnot A \lor \lnot B),$&$ (\lnot A \land \lnot B) $
        \\
        \quad \quad \quad $\left(Y_{1} \lor Y_{2}\right) \rightarrow \lnot Y_{3}$ $\rightsquigarrow$ $\left(Y_{1} \land Y_{2}\right) \rightarrow \lnot Y_{3}$ & 17 & 5.09\% & 6 & 9.84\% 
        & $ (A \lor B) \to \lnot C,$&$ (A \land B) \to \lnot C $
        \\
    \end{tabular}%
\end{table*}

\begin{table}
    \small
    \caption{Overview of the data sets, number of clusters found by our workflow, and the percentage of pairs explained by clusters for the propositional logic (PL), modal logic (PL) and computation tree logic (CTL) data sets. For each data set, we provide the number of (incorrect, correct) formula pairs, modelling contexts (i.e.\ different assignments), and statements to be modelled.}\label{tab:overview-clusters-explained}
    \begin{tabular}{l | r r r}
        &  \multicolumn{1}{c}{PL} & \multicolumn{1}{c}{ML} & \multicolumn{1}{c}{CTL} \\
        \midrule
        \# total formula pairs & $ 6106 $ & $ 12482 $ & $ 7210 $ \\
        \# modelling contexts & $ 11 $ & $ 6 $ & $ 6 $ \\
        \# statements to be modelled & $ 58 $ & $ 28 $ & $ 20 $ \\
        \hline
        \# total clusters & 248 & 1585 & 271 \\
        \% total explained & $ 84.44 $\% & $ 79.39 $\% & $ 35.89 $\% \\
        \hline
        \# clusters explaining $ \ge 0.1 $\% & 78 & 132 & 62\\
        \% explained by clusters explaining $ \ge 0.1 $\% & $ 76.06 $\% & $ 56.00 $\% & $ 29.76 $\% \\
        \hline
        \# clusters explaining $ \ge 0.5 $\% & 21 & 38 & 16\\
        \% explained by clusters explaining $ \ge 0.5 $\% & $ 66.34 $\% & $ 40.36 $\% & $ 19.78 $\% \\
        \hline
        \# clusters explaining $ \ge 1 $\% & 15 & 17 & 8 \\
        \% explained by clusters explaining $ \ge 1 $\% & $ 62.48 $\% & $ 28.26 $\% & $ 13.86 $\% 
    \end{tabular}%
\end{table}

We conducted two different experiments. First, we compared hand-picked mistakes to transformations found by our workflow as follows: For each hand-picked mistake, we identified all transformations that explain at least one pair that is explained by the hand-picked mistake and clustered them as described in \cref{section:clustering}.
As an example, \cref{tab:PropositionalUngroupedSorted-errors-trafos_main} shows a hand-picked mistake that a disjunction was erroneously used instead of a conjunction, and the transformations corresponding to it. This particular hand-picked mistake occurred in 334 pairs (column \#), which is 5.47\% of the whole data set. Of these, 330 (98.80\% of 334) were explained by bug fixing transformations.

Across the whole data set, the bug fixing transformations found by our workflow explain 5156 of the pairs ($ 84.44 $\%), whereas the hand-picked mistakes only explain 4370 of the pairs ($ 71.57 $\%). 918 pairs ($ 15.03 $\%) are exclusively explained by the transformations, i.e. no hand-picked mistake explains any of these pairs; 132 pairs ($ 2.16 $\%) exclusively by the hand-picked mistakes.
For instance, the following bug fixing transformations exclusively explain pairs that have not been explained by common mistakes in previous work

\begin{center}
    \midlabelled{$ \rho_{\text{xor-nand}} $}{
    \scalebox{.7}{%
        \trafo{
            \tree{%
                \Tree[.\ensuremath{\lor} 
                    [.\ensuremath{\land} 
                        \ensuremath{Y_1} 
                        [.\ensuremath{\lnot} \ensuremath{Y_2} ] 
                    ] 
                    [.\ensuremath{\land} 
                        \ensuremath{Y_2} 
                        [.\ensuremath{\lnot} \ensuremath{Y_1} ] 
                    ] 
                ]
            }
        }{
            \tree{%
                \Tree[.\ensuremath{\lnot} [.\ensuremath{\land} \ensuremath{Y_1} \ensuremath{Y_2} ] ]
            }
        }
    }}, and
    \midlabelled{$ \rho_{\text{at-least-2}} $}{
    \scalebox{.7}{%
        \trafo{
            \tree{%
                \Tree[.\ensuremath{\lor} \ensuremath{Y_1} \ensuremath{Y_2} \ensuremath{Y_3} ]
            }
        }{
            \tree{%
                \Tree[.\ensuremath{\lor} 
                    [.\ensuremath{\land} 
                        \ensuremath{Y_1} 
                        \ensuremath{Y_2}
                    ] 
                    [.\ensuremath{\land} 
                        \ensuremath{Y_1} 
                        \ensuremath{Y_3}
                    ] 
                    [.\ensuremath{\land} 
                        \ensuremath{Y_2} 
                        \ensuremath{Y_3}
                    ] 
                ]
            }
        }
    }}.
\end{center}

The transformation $ \rho_{\text{xor-nand}} $ represents the common mistake of using an exclusive or, expressed using a disjunction of conjunctions, instead of a negated conjunction which explains 114 pairs ($ 6.57 $\% of the pairs unexplained by the hand-picked mistakes). 
While a similar hand-picked mistake was found in \cite{SchmellenkampLZ23}, theirs only considers the case in which the exclusive or is expressed using a bi-implication, and thus does not explain the pairs explained by $ \rho_{\text{xor-nand}} $.

The transformation $ \rho_{\text{at-least-2}} $, which represents the common mistake of only using a disjunction to express that at least two literals are true, explains 28 pairs ($ 1.61 $\% of the pairs unexplained by the hand-picked mistakes).

\subsection{Results for Modal Logic and CTL}

The data set for modal logic contains $ 12482 $ pairs with $ 3951 $ of them being distinct and is also from \cite{SchmellenkampLZ23}. The data set for CTL contains $ 7210 $ pairs with $ 2695 $ distinct pairs and was collected by us using an interactive learning tool in two iterations of an introductory logic course (summer 2025 and 2026). The results for the modal logic and CTL data sets are summarized in \cref{tab:overview-clusters-explained}.

For modal logic, the transformations found by our workflow explain $ 79.39 $\% of the pairs in the data set, which is comparable to the propositional logic case.  
Examples of transformations are

\begin{center}
\midlabelled{$ \rho^\text{ml}_1 $}{
    \hspace*{-1em}
    \scalebox{.7}{%
    \trafo{
        \tree{%
            \Tree[.\ensuremath{\Box}
                \ensuremath{Y_1}
            ]
        }
    }{
        \tree{%
            \Tree[.\ensuremath{\Diamond}
                \ensuremath{Y_1}
            ]
        }
    }
}}\hspace*{-0.5em},
\midlabelled{$ \rho^\text{ml}_2 $}{
    \hspace*{-1em}
    \scalebox{.7}{%
    \trafo{
        \tree{%
            \Tree[.\ensuremath{Y_1} ]
        }
    }{
        \tree{%
            \Tree[.\ensuremath{\Box}
                \ensuremath{Y_1}
            ]
        }
    }
}}\hspace*{-0.5em}, and
\midlabelled{$ \rho^\text{ml}_3 $}{
    \hspace*{-1em}
    \scalebox{.7}{%
    \trafo{
        \tree{%
            \Tree[.\ensuremath{\lor}
                \ensuremath{Y_1}
                \ensuremath{Y_2}
            ]
        }
    }{
        \tree{%
            \Tree[.\ensuremath{\lnot}
                [.\ensuremath{\leftrightarrow}
                    \ensuremath{Y_1}
                    \ensuremath{Y_2}
                ]
            ]
        }
    }
}}\hspace*{-0.5em},
\end{center}

which explain $ 2.1 $\%, $ 1.43 $\%, and $ 3.37 $\% of the data.
While $ \rho^\text{ml}_1 $ and $ \rho^\text{ml}_2 $ change or add the modal operators $ \Diamond $ and $ \Box $,
$ \rho^\text{modal}_3 $ was also found for the propositional logic data set. It represents the common mistake of modelling an ``exclusive or'' as an ``or''.

For CTL, the transformations found by our workflow explain $ 35.89 $\% of the pairs with the two transformations that explain the largest fractions of the data set being

\begin{center}
\midlabelled{$ \rho^\text{ctl}_1 $}{
\scalebox{.7}{%
    \trafo{
        \tree{%
            \Tree[.\ensuremath{Y_1} ]
        }
    }{
        \tree{%
            \Tree[.\ensuremath{\mathbf{A} \mathbf{G}}
                \ensuremath{Y_1}
            ]
        }
    }
}}, and
\midlabelled{$ \rho^\text{ctl}_2 $}{
\scalebox{.7}{%
    \trafo{
        \tree{%
            \Tree[.\ensuremath{\mathbf{A} \mathbf{X}}
                \ensuremath{Y_1}
            ]
        }
    }{
        \tree{%
            \Tree[.\ensuremath{\mathbf{A} \mathbf{G}}
                \ensuremath{Y_1}
            ]
        }
    }
}}
\end{center}

explaining $ 2.98 $\% and $ 2.48 $\%, respectively.

\subsubsection*{Determining Usefulness of Transformations}
To exemplarily determine the usefulness of clusters of transformations found by our workflow, a CS educator inspected all clusters explaining at least $ 0.5 $\% of the pairs in each of the three data sets for propositional logic, modal logic, and CTL (see also \cref{tab:overview-clusters-explained}).
The focus was on whether the ``main transformation'', i.e. the transformation forming the root of the cluster tree, is sufficient to provide targeted feedback, or if sub-transformations, those lower in the cluster's hierarchy, would be required for this.

In propositional logic, 19 clusters had main transformations suitable for feedback, though at least 4 required sub-transformations for more targeted feedback. In CTL, all 17 clusters had useful main transformations, with at least 6 requiring sub-transformations for sufficiently specific feedback. In modal logic, the main transformation of 18 clusters was suitable for feedback, but
at least 7 required sub-transformations. In 16 clusters, the main transformations were too general to identify specific mistakes directly; for 5 of these, frequent sub-transformations revealed mistakes.

\section{Discussion and Conclusion}

We presented a tool-supported workflow for identifying candidates for common mistakes in modelling with mathematical formalisms as bug fixing transformations, clustering them, and presenting them to educators and CS education researchers. Our experiments
show that the workflow recovers many known propositional logic mistakes, finds
additional ones, and can also be applied to other formalisms, such as modal logic and CTL.

Beyond modelling, the workflow is likely also applicable to tasks in which students
transform logical formulas into normal forms. A difference in that setting is that 
one does not immediately have (incorrect, correct) pairs. Instead, one has an incorrect formula, a previous correct formula, and the
equivalence transformation that was applied incorrectly. A suitable pair can then
be constructed by applying the intended equivalence transformation to the
previous correct formula, yielding the correct formula for that step.

Overall, the results suggest that transformation-based mistake candidates can
help scale the discovery of common mistakes in education settings. 
Future work includes applying the workflow to diverse data sets from different domains, 
and studying how the resulting candidates can be integrated into feedback systems of interactive learning tools.

\begin{acks}
    This work was supported by the Deutsche Forschungsgemeinschaft (DFG, German Research Foundation), grant 448468041.
\end{acks}

\bibliographystyle{ACM-Reference-Format}
\bibliography{bibliography}

\clearpage
\appendix

\onecolumn
\section{Prompts for Finding Bug Fixing Transformations}\label{section:appendix-prompts}
The prompts for all propositional logic, modal logic, and CTL share large parts; they mainly differ in the symbols the transformations are allowed to use. 
The following prompt represents all three prompts, with the only differing sentence being prefixed by the logic the prompt is used for.

\subsection*{Start of prompt:}

\subsubsection*{Instruction}
Find structural differences between pairs of propositional logic formulas and express them by transformations of a transformation language.
A transformation of our transformation language consists of two tree patterns:
\begin{itemize}
    \item \texttt{body}: the pattern that matches a subtree in the input tree (left side of the transformation).
    \item \texttt{head}: the pattern that replaces the matched subtree (right side of the transformation).
    \item Transformations are written in the format: \texttt{body} ---> \texttt{head}
\end{itemize}%

Patterns are trees that follow the following rules:
\begin{itemize}
    \item 
        \adjustbox{valign=t}{\begin{tabularx}{\textwidth-2em}{l X}
            \emph{propositional}: & Inner nodes are labeled by the operators of propositional logic: ``or'', ``and'', ``impl'', ``not'', ``eq''.\\
            \emph{modal}: & Inner nodes are labeled by the operators of modal logic: ``or'', ``and'', ``impl'', ``not'', ``eq'', ``box'' (for $ \Box $), ``dia'' (for $ \Diamond $).\\
            \emph{CTL}: & Inner nodes are labeled by the operators of computation tree logic: ``or'', ``and'', ``impl'', ``not'', ``eq'', ``ax'', ``ex'', ``af'', ``ef'', ``ag'', ``eg'', ``au'', ``eu''.
        \end{tabularx}}
        
    \item They are written as: operator[child1 child2 ...] with children separated by spaces.
    \item Leaf nodes are labeled by tree variables, which are placeholders for subtrees. Write them as Y1, Y2, ... 
    \item Variadic patterns: ``and'' and ``or'' may have arbitrary arity. Use Y1 .. Yn to denote an ordered sequence of two or more child subtrees.\\
        Example: and[Y1 .. Yn] matches and[A B], and[A B C], and[A B C D], etc.; or[Y1 .. Yn] matches or[A B], or[A B C], or[A B C D], etc.
    \item The sequence Y1 .. Yn must be preserved unchanged: do not reorder, duplicate, or split it. Do not remove or add single elements.
    \item Use Y1 .. Yn only as the complete child list of an and/or node. \\
        Example: and[Y1 .. Yn] 
    \item Every tree variable that appears in the \texttt{head} must also appear in the \texttt{body}.
    \item Propositional variables of the formulas may not appear in the transformations. Use only tree variables and the operators as node labels.
\end{itemize}%

\subsubsection*{Matching:}
\begin{itemize}
    \item A match for the \texttt{body} may occur anywhere in the tree, not only at the root.
    \item Matching is injective: every \texttt{body}-node is mapped to a unique node in the input tree.
    \item If tree variables appear multiple times in the same pattern, the occurrences have to be matched by isomorphic subtrees.
\end{itemize}%

\subsubsection*{Application:}
When a match is found, the corresponding subtree is replaced by the head,
with variables substituted by the subtrees/labels found in the match.

\subsubsection*{Your Task:}
Find \texttt{<n\_transformations>} transformation(s) that can transform as many of the formula pairs as possible in one step.
Return the answer exclusively in the following JSON format. Write nothing outside this JSON. Pay attention to correct quotation marks, no comments, no headings. Replace TRANSFORMATION by the respective transformation.

\begin{lstlisting}
{
  "transformations": [
    {
      "name": "Transformation_1",
      "transformation": "TRANSFORMATION"
    }
  ]
}
\end{lstlisting}

\noindent
Here is the dataset of formula pairs: 

\begin{lstlisting}
Input: <error_formula_1> Target: <correct_formula_1>
Input: <error_formula_2> Target: <correct_formula_2>
\end{lstlisting}
\vdots
 
\clearpage
\section{Detailed Evaluation Data}

\subsection{Propositional Logic Data Set Grouped by Linguistic Operator}
The hand-picked mistakes are from \cite{SchmellenkampLZ23}.
\begin{longtable}{lrrrr}
\caption{Hand-picked errors and transformations for linguistic operator(s) AND\_AND\_NONE\_OF}\label{tab:AND_AND_NONE_OF-errors-trafos}\\
\toprule
Hand-picked mistake / Transformations & \# & \% & \#dst & \%dst \\
\midrule
\endfirsthead
\toprule
Hand-picked mistake / Transformations & \# & \% & \#dst & \%dst \\
\midrule
\endhead
\midrule
\multicolumn{5}{r}{Continued on next page} \\
\endfoot
\bottomrule
\endlastfoot
\midrule
\textbf{Unexplained by ALL hand errors} & \textbf{48} & \textbf{100.00\%} & \textbf{26} & \textbf{100.00\%} \\
\textbf{\quad Transformations combined} & \textbf{13} & \textbf{27.08\%} & \textbf{12} & \textbf{46.15\%} \\
\quad T1: $Y_{1} \land Y_{2} \land \lnot \left(Y_{3} \land Y_{4}\right)$ $\rightsquigarrow$ $Y_{1} \land Y_{2} \land \lnot Y_{3} \land \lnot Y_{4}$ & 4 & 8.33\% & 3 & 11.54\% \\
\quad T2: $\left(Y_{1} \land Y_{2}\right) \leftrightarrow \lnot \left(Y_{3} \land Y_{4}\right)$ $\rightsquigarrow$ $Y_{1} \land Y_{2} \land \lnot Y_{3} \land \lnot Y_{4}$ & 1 & 2.08\% & 1 & 3.85\% \\
\quad T3: $\left(Y_{1} \land Y_{2}\right) \rightarrow \left(\lnot Y_{3} \land \lnot Y_{4}\right)$ $\rightsquigarrow$ $Y_{1} \land Y_{2} \land \lnot Y_{3} \land \lnot Y_{4}$ & 1 & 2.08\% & 1 & 3.85\% \\
\quad T4: $\left(Y_{1} \land Y_{2}\right) \leftrightarrow \left(\lnot Y_{3} \land \lnot Y_{4}\right)$ $\rightsquigarrow$ $Y_{1} \land Y_{2} \land \lnot Y_{3} \land \lnot Y_{4}$ & 1 & 2.08\% & 1 & 3.85\% \\
\quad T5: $\left(Y_{1} \land Y_{2} \land \lnot Y_{3}\right) \lor \left(Y_{1} \land Y_{2} \land \lnot Y_{4}\right)$ $\rightsquigarrow$ $Y_{1} \land Y_{2} \land \lnot Y_{3} \land \lnot Y_{4}$ & 1 & 2.08\% & 1 & 3.85\% \\
\quad T6: $\left(\lnot Y_{1} \land \lnot Y_{2}\right) \rightarrow \left(Y_{3} \land Y_{4}\right)$ $\rightsquigarrow$ $Y_{3} \land Y_{4} \land \lnot Y_{1} \land \lnot Y_{2}$ & 1 & 2.08\% & 1 & 3.85\% \\
\end{longtable}

\clearpage

\begin{longtable}{lrrrr}
\caption{Hand-picked errors and transformations for linguistic operator(s) AND\_AND}\label{tab:AND_AND-errors-trafos}\\
\toprule
Hand-picked mistake / Transformations & \# & \% & \#dst & \%dst \\
\midrule
\endfirsthead
\toprule
Hand-picked mistake / Transformations & \# & \% & \#dst & \%dst \\
\midrule
\endhead
\midrule
\multicolumn{5}{r}{Continued on next page} \\
\endfoot
\bottomrule
\endlastfoot
\midrule
\textbf{BUGGY\_ADD\_OPERAND\_TO\_CONJUNCTION} & \textbf{19} & \textbf{65.52\%} & \textbf{4} & \textbf{30.77\%} \\
\textbf{\quad Transformations combined} & \textbf{0} & \textbf{0.00\%} & \textbf{0} & \textbf{0.00\%} \\
\midrule
\textbf{BUGGY\_REPLACE\_DISJUNCTION\_BY\_CONJUNCTION} & \textbf{2} & \textbf{6.90\%} & \textbf{2} & \textbf{15.38\%} \\
\textbf{\quad Transformations combined} & \textbf{2} & \textbf{100.00\%} & \textbf{2} & \textbf{100.00\%} \\
\quad T1: $Y_{1} \lor \ldots \lor Y_{n}$ $\rightsquigarrow$ $Y_{1} \land \ldots \land Y_{n}$ & 1 & 50.00\% & 1 & 50.00\% \\
\quad T4: $\left(Y_{1} \lor Y_{2}\right) \land Y_{3}$ $\rightsquigarrow$ $Y_{3} \land Y_{1} \land Y_{2}$ & 1 & 50.00\% & 1 & 50.00\% \\
\midrule
\textbf{BUGGY\_REPLACE\_EQUIVALENCE\_BY\_CONJUNCTION} & \textbf{2} & \textbf{6.90\%} & \textbf{1} & \textbf{7.69\%} \\
\textbf{\quad Transformations combined} & \textbf{2} & \textbf{100.00\%} & \textbf{1} & \textbf{100.00\%} \\
\end{longtable}

\clearpage

\begin{longtable}{lrrrr}
\caption{Hand-picked errors and transformations for linguistic operator(s) AND\_IF}\label{tab:AND_IF-errors-trafos}\\
\toprule
Hand-picked mistake / Transformations & \# & \% & \#dst & \%dst \\
\midrule
\endfirsthead
\toprule
Hand-picked mistake / Transformations & \# & \% & \#dst & \%dst \\
\midrule
\endhead
\midrule
\multicolumn{5}{r}{Continued on next page} \\
\endfoot
\bottomrule
\endlastfoot
\midrule
\textbf{BUGGY\_REPLACE\_IMPLICATION\_BY\_REVERSED\_IMPLICATION} & \textbf{12} & \textbf{8.89\%} & \textbf{1} & \textbf{1.69\%} \\
\textbf{\quad Transformations combined} & \textbf{12} & \textbf{100.00\%} & \textbf{1} & \textbf{100.00\%} \\
\quad T2: $\left(Y_{1} \rightarrow Y_{2}\right) \land Y_{3}$ $\rightsquigarrow$ $\left(Y_{2} \rightarrow Y_{1}\right) \land Y_{3}$ & 12 & 100.00\% & 1 & 100.00\% \\
\midrule
\textbf{BUGGY\_REPLACE\_IMPLICATION\_BY\_CONJUNCTION} & \textbf{10} & \textbf{7.41\%} & \textbf{3} & \textbf{5.08\%} \\
\textbf{\quad Transformations combined} & \textbf{8} & \textbf{80.00\%} & \textbf{1} & \textbf{33.33\%} \\
\quad T4: $Y_{1} \rightarrow \left(Y_{2} \rightarrow Y_{3}\right)$ $\rightsquigarrow$ $\left(Y_{2} \rightarrow Y_{3}\right) \land Y_{1}$ & 8 & 80.00\% & 1 & 33.33\% \\
\midrule
\textbf{BUGGY\_REPLACE\_EQUIVALENCE\_BY\_IMPLICATION} & \textbf{9} & \textbf{6.67\%} & \textbf{1} & \textbf{1.69\%} \\
\textbf{\quad Transformations combined} & \textbf{9} & \textbf{100.00\%} & \textbf{1} & \textbf{100.00\%} \\
\end{longtable}

\clearpage

\begin{longtable}{lrrrr}
\caption{Hand-picked errors and transformations for linguistic operator(s) AND\_NOT\_NOT}\label{tab:AND_NOT_NOT-errors-trafos}\\
\toprule
Hand-picked mistake / Transformations & \# & \% & \#dst & \%dst \\
\midrule
\endfirsthead
\toprule
Hand-picked mistake / Transformations & \# & \% & \#dst & \%dst \\
\midrule
\endhead
\midrule
\multicolumn{5}{r}{Continued on next page} \\
\endfoot
\bottomrule
\endlastfoot
\midrule
\textbf{BUGGY\_REPLACE\_EQUIVALENCE\_BY\_CONJUNCTION} & \textbf{1} & \textbf{100.00\%} & \textbf{1} & \textbf{100.00\%} \\
\textbf{\quad Transformations combined} & \textbf{1} & \textbf{100.00\%} & \textbf{1} & \textbf{100.00\%} \\
\quad T1: $Y_{1} \leftrightarrow Y_{2}$ $\rightsquigarrow$ $Y_{1} \land Y_{2}$ & 1 & 100.00\% & 1 & 100.00\% \\
\midrule
\textbf{Unexplained by ALL hand errors} & \textbf{0} & \textbf{0.00\%} & \textbf{0} & \textbf{0.00\%} \\
\textbf{\quad Transformations combined} & \textbf{0} & \textbf{0.00\%} & \textbf{0} & \textbf{0.00\%} \\
\midrule
Whole file & 1 & 100.00\% & 1 & 100.00\% \\
\textbf{ANY hand error explained} & \textbf{1} & \textbf{100.00\%} & \textbf{1} & \textbf{100.00\%} \\
\textbf{ANY transformation explained} & \textbf{1} & \textbf{100.00\%} & \textbf{1} & \textbf{100.00\%} \\
Only explained by hand error & 0 & 0.00\% & 0 & 0.00\% \\
Only explained by transformation & 0 & 0.00\% & 0 & 0.00\% \\
\end{longtable}

\clearpage

\begin{longtable}{lrrrr}
\caption{Hand-picked errors and transformations for linguistic operator(s) AND\_NOT}\label{tab:AND_NOT-errors-trafos}\\
\toprule
Hand-picked mistake / Transformations & \# & \% & \#dst & \%dst \\
\midrule
\endfirsthead
\toprule
Hand-picked mistake / Transformations & \# & \% & \#dst & \%dst \\
\midrule
\endhead
\midrule
\multicolumn{5}{r}{Continued on next page} \\
\endfoot
\bottomrule
\endlastfoot
\midrule
\textbf{BUGGY\_REPLACE\_DISJUNCTION\_BY\_CONJUNCTION} & \textbf{4} & \textbf{80.00\%} & \textbf{1} & \textbf{50.00\%} \\
\textbf{\quad Transformations combined} & \textbf{4} & \textbf{100.00\%} & \textbf{1} & \textbf{100.00\%} \\
\quad T1: $Y_{1} \lor \lnot Y_{2}$ $\rightsquigarrow$ $Y_{1} \land \lnot Y_{2}$ & 4 & 100.00\% & 1 & 100.00\% \\
\midrule
\textbf{BUGGY\_REPLACE\_IMPLICATION\_BY\_CONJUNCTION} & \textbf{1} & \textbf{20.00\%} & \textbf{1} & \textbf{50.00\%} \\
\textbf{\quad Transformations combined} & \textbf{1} & \textbf{100.00\%} & \textbf{1} & \textbf{100.00\%} \\
\quad T2: $Y_{1} \rightarrow Y_{2}$ $\rightsquigarrow$ $Y_{1} \land Y_{2}$ & 1 & 100.00\% & 1 & 100.00\% \\
\midrule
\textbf{Unexplained by ALL hand errors} & \textbf{0} & \textbf{0.00\%} & \textbf{0} & \textbf{0.00\%} \\
\textbf{\quad Transformations combined} & \textbf{0} & \textbf{0.00\%} & \textbf{0} & \textbf{0.00\%} \\
\midrule
Whole file & 5 & 100.00\% & 2 & 100.00\% \\
\textbf{ANY hand error explained} & \textbf{5} & \textbf{100.00\%} & \textbf{2} & \textbf{100.00\%} \\
\textbf{ANY transformation explained} & \textbf{5} & \textbf{100.00\%} & \textbf{2} & \textbf{100.00\%} \\
Only explained by hand error & 0 & 0.00\% & 0 & 0.00\% \\
Only explained by transformation & 0 & 0.00\% & 0 & 0.00\% \\
\end{longtable}

\clearpage

\begin{longtable}{lrrrr}
\caption{Hand-picked errors and transformations for linguistic operator(s) AND}\label{tab:AND-errors-trafos}\\
\toprule
Hand-picked mistake / Transformations & \# & \% & \#dst & \%dst \\
\midrule
\endfirsthead
\toprule
Hand-picked mistake / Transformations & \# & \% & \#dst & \%dst \\
\midrule
\endhead
\midrule
\multicolumn{5}{r}{Continued on next page} \\
\endfoot
\bottomrule
\endlastfoot
\midrule
\textbf{BUGGY\_REPLACE\_DISJUNCTION\_BY\_CONJUNCTION} & \textbf{50} & \textbf{78.12\%} & \textbf{6} & \textbf{40.00\%} \\
\textbf{\quad Transformations combined} & \textbf{50} & \textbf{100.00\%} & \textbf{6} & \textbf{100.00\%} \\
\quad $\lceil$ $=$ T1: $Y_{1} \lor \ldots \lor Y_{n}$ $\rightsquigarrow$ $Y_{1} \land \ldots \land Y_{n}$ & 50 & 100.00\% & 6 & 100.00\% \\
\quad $\lfloor$ $=$ T2: $Y_{1} \lor Y_{2}$ $\rightsquigarrow$ $Y_{1} \land Y_{2}$ & 50 & 100.00\% & 6 & 100.00\% \\
\midrule
\textbf{BUGGY\_REPLACE\_IMPLICATION\_BY\_CONJUNCTION} & \textbf{7} & \textbf{10.94\%} & \textbf{3} & \textbf{20.00\%} \\
\textbf{\quad Transformations combined} & \textbf{7} & \textbf{100.00\%} & \textbf{3} & \textbf{100.00\%} \\
\quad T3: $Y_{1} \rightarrow Y_{2}$ $\rightsquigarrow$ $Y_{2} \land Y_{1}$ & 6 & 85.71\% & 2 & 66.67\% \\
\quad T5: $Y_{1} \rightarrow Y_{2}$ $\rightsquigarrow$ $Y_{1} \land Y_{2}$ & 1 & 14.29\% & 1 & 33.33\% \\
\end{longtable}

\clearpage

\begin{longtable}{p{0.58\textwidth} rrrr}
\caption{Hand-picked errors and transformations for linguistic operator(s) AT\_LEAST\_NEUTRAL}\label{tab:AT_LEAST_NEUTRAL-errors-trafos}\\
\toprule
Hand-picked mistake / Transformations & \# & \% & \#dst & \%dst \\
\midrule
\endfirsthead
\toprule
Hand-picked mistake / Transformations & \# & \% & \#dst & \%dst \\
\midrule
\endhead
\midrule
\multicolumn{5}{r}{Continued on next page} \\
\endfoot
\bottomrule
\endlastfoot
\midrule
\textbf{BUGGY\_REPLACE\_NAND\_BY\_DISJUNCTION} & \textbf{2} & \textbf{40.00\%} & \textbf{1} & \textbf{25.00\%} \\
\textbf{\quad Transformations combined} & \textbf{2} & \textbf{100.00\%} & \textbf{1} & \textbf{100.00\%} \\
\quad T1: $\lnot \left(Y_{1} \land Y_{2} \land Y_{3} \land Y_{4}\right)$ $\rightsquigarrow$ $Y_{1} \lor Y_{2} \lor Y_{3} \lor Y_{4}$ & 2 & 100.00\% & 1 & 100.00\% \\
\midrule
\textbf{BUGGY\_ADD\_OPERAND\_TO\_DISJUNCTION} & \textbf{1} & \textbf{20.00\%} & \textbf{1} & \textbf{25.00\%} \\
\textbf{\quad Transformations combined} & \textbf{0} & \textbf{0.00\%} & \textbf{0} & \textbf{0.00\%} \\
\midrule
\textbf{Unexplained by ALL hand errors} & \textbf{2} & \textbf{40.00\%} & \textbf{2} & \textbf{50.00\%} \\
\textbf{\quad Transformations combined} & \textbf{2} & \textbf{100.00\%} & \textbf{2} & \textbf{100.00\%} \\
\quad T2: $\left(Y_{1} \land \lnot Y_{2} \land \lnot Y_{3} \land \lnot Y_{4}\right) \lor \left(Y_{2} \land \lnot Y_{1} \land \lnot Y_{3} \land \lnot Y_{4}\right) \lor \left(Y_{3} \land \lnot Y_{1} \land \lnot Y_{2} \land \lnot Y_{4}\right) \lor \left(Y_{4} \land \lnot Y_{1} \land \lnot Y_{2} \land \lnot Y_{3}\right)$ $\rightsquigarrow$ $Y_{1} \lor Y_{2} \lor Y_{3} \lor Y_{4}$ & 1 & 50.00\% & 1 & 50.00\% \\
\end{longtable}

\clearpage

\begin{longtable}{lrrrr}
\caption{Hand-picked errors and transformations for linguistic operator(s) AT\_LEAST}\label{tab:AT_LEAST-errors-trafos}\\
\toprule
Hand-picked mistake / Transformations & \# & \% & \#dst & \%dst \\
\midrule
\endfirsthead
\toprule
Hand-picked mistake / Transformations & \# & \% & \#dst & \%dst \\
\midrule
\endhead
\midrule
\multicolumn{5}{r}{Continued on next page} \\
\endfoot
\bottomrule
\endlastfoot
\midrule
\textbf{BUGGY\_REPLACE\_DISJUNCTION\_BY\_CONJUNCTION} & \textbf{3} & \textbf{1.46\%} & \textbf{2} & \textbf{1.61\%} \\
\textbf{\quad Transformations combined} & \textbf{3} & \textbf{100.00\%} & \textbf{2} & \textbf{100.00\%} \\
\quad T10: $\left(Y_{1} \land Y_{2}\right) \lor \left(Y_{1} \land Y_{3}\right) \lor Y_{2} \lor Y_{3}$ $\rightsquigarrow$ $\left(Y_{1} \land Y_{2}\right) \lor \left(Y_{1} \land Y_{3}\right) \lor \left(Y_{2} \land Y_{3}\right)$ & 3 & 100.00\% & 2 & 100.00\% \\
\midrule
\textbf{BUGGY\_REPLACE\_CONJUNCTION\_BY\_DISJUNCTION} & \textbf{2} & \textbf{0.97\%} & \textbf{2} & \textbf{1.61\%} \\
\textbf{\quad Transformations combined} & \textbf{0} & \textbf{0.00\%} & \textbf{0} & \textbf{0.00\%} \\
\midrule
\textbf{BUGGY\_REPLACE\_NAND\_BY\_DISJUNCTION} & \textbf{2} & \textbf{0.97\%} & \textbf{2} & \textbf{1.61\%} \\
\textbf{\quad Transformations combined} & \textbf{0} & \textbf{0.00\%} & \textbf{0} & \textbf{0.00\%} \\
\midrule
\textbf{Unexplained by ALL hand errors} & \textbf{199} & \textbf{96.60\%} & \textbf{118} & \textbf{95.16\%} \\
\end{longtable}

\clearpage

\begin{longtable}{lrrrr}
\caption{Hand-picked errors and transformations for linguistic operator(s) EITHER\_OR}\label{tab:EITHER_OR-errors-trafos}\\
\toprule
Hand-picked mistake / Transformations & \# & \% & \#dst & \%dst \\
\midrule
\endfirsthead
\toprule
Hand-picked mistake / Transformations & \# & \% & \#dst & \%dst \\
\midrule
\endhead
\midrule
\multicolumn{5}{r}{Continued on next page} \\
\endfoot
\bottomrule
\endlastfoot
\midrule
\textbf{BUGGY\_REPLACE\_DISJUNCTION\_BY\_XOR} & \textbf{91} & \textbf{50.00\%} & \textbf{2} & \textbf{6.25\%} \\
\textbf{\quad Transformations combined} & \textbf{91} & \textbf{100.00\%} & \textbf{2} & \textbf{100.00\%} \\
\quad T1: $Y_{1} \lor Y_{2}$ $\rightsquigarrow$ $\lnot \left(Y_{1} \leftrightarrow Y_{2}\right)$ & 91 & 100.00\% & 2 & 100.00\% \\
\midrule
\textbf{BUGGY\_REPLACE\_EQUIVALENCE\_BY\_XOR} & \textbf{21} & \textbf{11.54\%} & \textbf{1} & \textbf{3.12\%} \\
\textbf{\quad Transformations combined} & \textbf{21} & \textbf{100.00\%} & \textbf{1} & \textbf{100.00\%} \\
\quad T2: $Y_{1} \leftrightarrow Y_{2}$ $\rightsquigarrow$ $\lnot \left(Y_{1} \leftrightarrow Y_{2}\right)$ & 21 & 100.00\% & 1 & 100.00\% \\
\midrule
\textbf{BUGGY\_REPLACE\_NAND\_BY\_XOR} & \textbf{12} & \textbf{6.59\%} & \textbf{3} & \textbf{9.38\%} \\
\textbf{\quad Transformations combined} & \textbf{11} & \textbf{91.67\%} & \textbf{2} & \textbf{66.67\%} \\
\end{longtable}

\clearpage

\begin{longtable}{lrrrr}
\caption{Hand-picked errors and transformations for linguistic operator(s) IF\_AND\_NEUTRAL}\label{tab:IF_AND_NEUTRAL-errors-trafos}\\
\toprule
Hand-picked mistake / Transformations & \# & \% & \#dst & \%dst \\
\midrule
\endfirsthead
\toprule
Hand-picked mistake / Transformations & \# & \% & \#dst & \%dst \\
\midrule
\endhead
\midrule
\multicolumn{5}{r}{Continued on next page} \\
\endfoot
\bottomrule
\endlastfoot
\midrule
\textbf{BUGGY\_REMOVE\_NEGATION} & \textbf{9} & \textbf{37.50\%} & \textbf{3} & \textbf{21.43\%} \\
\textbf{\quad Transformations combined} & \textbf{9} & \textbf{100.00\%} & \textbf{3} & \textbf{100.00\%} \\
\quad T4: $Y_{1} \rightarrow \lnot Y_{2}$ $\rightsquigarrow$ $Y_{1} \rightarrow Y_{2}$ & 9 & 100.00\% & 3 & 100.00\% \\
\midrule
\textbf{BUGGY\_REPLACE\_IMPLICATION\_BY\_REVERSED\_IMPLICATION} & \textbf{4} & \textbf{16.67\%} & \textbf{2} & \textbf{14.29\%} \\
\textbf{\quad Transformations combined} & \textbf{4} & \textbf{100.00\%} & \textbf{2} & \textbf{100.00\%} \\
\quad T1: $Y_{1} \rightarrow Y_{2}$ $\rightsquigarrow$ $Y_{2} \rightarrow Y_{1}$ & 4 & 100.00\% & 2 & 100.00\% \\
\midrule
\textbf{BUGGY\_REPLACE\_DISJUNCTION\_BY\_CONJUNCTION} & \textbf{2} & \textbf{8.33\%} & \textbf{1} & \textbf{7.14\%} \\
\textbf{\quad Transformations combined} & \textbf{2} & \textbf{100.00\%} & \textbf{1} & \textbf{100.00\%} \\
\end{longtable}

\clearpage

\begin{longtable}{lrrrr}
\caption{Hand-picked errors and transformations for linguistic operator(s) IF\_AND\_NOT}\label{tab:IF_AND_NOT-errors-trafos}\\
\toprule
Hand-picked mistake / Transformations & \# & \% & \#dst & \%dst \\
\midrule
\endfirsthead
\toprule
Hand-picked mistake / Transformations & \# & \% & \#dst & \%dst \\
\midrule
\endhead
\midrule
\multicolumn{5}{r}{Continued on next page} \\
\endfoot
\bottomrule
\endlastfoot
\midrule
\textbf{BUGGY\_ADD\_NEGATION} & \textbf{24} & \textbf{33.80\%} & \textbf{6} & \textbf{15.00\%} \\
\textbf{\quad Transformations combined} & \textbf{24} & \textbf{100.00\%} & \textbf{6} & \textbf{100.00\%} \\
\quad T1: $\left(Y_{1} \land Y_{2}\right) \rightarrow Y_{3}$ $\rightsquigarrow$ $\left(Y_{1} \land Y_{2}\right) \rightarrow \lnot Y_{3}$ & 24 & 100.00\% & 6 & 100.00\% \\
\midrule
\textbf{BUGGY\_REPLACE\_IMPLICATION\_BY\_REVERSED\_IMPLICATION} & \textbf{12} & \textbf{16.90\%} & \textbf{5} & \textbf{12.50\%} \\
\textbf{\quad Transformations combined} & \textbf{12} & \textbf{100.00\%} & \textbf{5} & \textbf{100.00\%} \\
\quad T2: $\lnot Y_{1} \rightarrow Y_{2}$ $\rightsquigarrow$ $Y_{2} \rightarrow \lnot Y_{1}$ & 12 & 100.00\% & 5 & 100.00\% \\
\midrule
\textbf{BUGGY\_REPLACE\_DISJUNCTION\_BY\_CONJUNCTION} & \textbf{9} & \textbf{12.68\%} & \textbf{5} & \textbf{12.50\%} \\
\textbf{\quad Transformations combined} & \textbf{9} & \textbf{100.00\%} & \textbf{5} & \textbf{100.00\%} \\
\end{longtable}

\clearpage

\begin{longtable}{lrrrr}
\caption{Hand-picked errors and transformations for linguistic operator(s) IF\_AND\_ONLY\_IF}\label{tab:IF_AND_ONLY_IF-errors-trafos}\\
\toprule
Hand-picked mistake / Transformations & \# & \% & \#dst & \%dst \\
\midrule
\endfirsthead
\toprule
Hand-picked mistake / Transformations & \# & \% & \#dst & \%dst \\
\midrule
\endhead
\midrule
\multicolumn{5}{r}{Continued on next page} \\
\endfoot
\bottomrule
\endlastfoot
\midrule
\textbf{BUGGY\_REPLACE\_IMPLICATION\_BY\_EQUIVALENCE} & \textbf{628} & \textbf{88.20\%} & \textbf{39} & \textbf{52.00\%} \\
\textbf{\quad Transformations combined} & \textbf{626} & \textbf{99.68\%} & \textbf{38} & \textbf{97.44\%} \\
\quad T1: $Y_{1} \rightarrow Y_{2}$ $\rightsquigarrow$ $Y_{2} \leftrightarrow Y_{1}$ & 344 & 54.78\% & 20 & 51.28\% \\
\quad T2: $Y_{1} \rightarrow Y_{2}$ $\rightsquigarrow$ $Y_{1} \leftrightarrow Y_{2}$ & 282 & 44.90\% & 18 & 46.15\% \\
\midrule
\textbf{BUGGY\_REPLACE\_CONJUNCTION\_BY\_EQUIVALENCE} & \textbf{35} & \textbf{4.92\%} & \textbf{10} & \textbf{13.33\%} \\
\textbf{\quad Transformations combined} & \textbf{35} & \textbf{100.00\%} & \textbf{10} & \textbf{100.00\%} \\
\quad $\lceil$ $=$ T3: $Y_{1} \land Y_{2}$ $\rightsquigarrow$ $Y_{1} \leftrightarrow Y_{2}$ & 18 & 51.43\% & 4 & 40.00\% \\
\quad $\lfloor$ $=$ T4: $Y_{1} \land \ldots \land Y_{n}$ $\rightsquigarrow$ $Y_{1} \leftrightarrow Y_{2}$ & 18 & 51.43\% & 4 & 40.00\% \\
\end{longtable}

\clearpage

\begin{longtable}{lrrrr}
\caption{Hand-picked errors and transformations for linguistic operator(s) IF\_AND}\label{tab:IF_AND-errors-trafos}\\
\toprule
Hand-picked mistake / Transformations & \# & \% & \#dst & \%dst \\
\midrule
\endfirsthead
\toprule
Hand-picked mistake / Transformations & \# & \% & \#dst & \%dst \\
\midrule
\endhead
\midrule
\multicolumn{5}{r}{Continued on next page} \\
\endfoot
\bottomrule
\endlastfoot
\midrule
\textbf{BUGGY\_REPLACE\_IMPLICATION\_BY\_REVERSED\_IMPLICATION} & \textbf{6} & \textbf{25.00\%} & \textbf{3} & \textbf{17.65\%} \\
\textbf{\quad Transformations combined} & \textbf{6} & \textbf{100.00\%} & \textbf{3} & \textbf{100.00\%} \\
\quad T1: $Y_{1} \rightarrow Y_{2}$ $\rightsquigarrow$ $Y_{2} \rightarrow Y_{1}$ & 6 & 100.00\% & 3 & 100.00\% \\
\midrule
\textbf{BUGGY\_REPLACE\_EQUIVALENCE\_BY\_REVERSED\_IMPLICATION} & \textbf{2} & \textbf{8.33\%} & \textbf{2} & \textbf{11.76\%} \\
\textbf{\quad Transformations combined} & \textbf{2} & \textbf{100.00\%} & \textbf{2} & \textbf{100.00\%} \\
\quad T2: $Y_{1} \leftrightarrow Y_{2}$ $\rightsquigarrow$ $Y_{2} \rightarrow Y_{1}$ & 2 & 100.00\% & 2 & 100.00\% \\
\midrule
\textbf{BUGGY\_REPLACE\_CONJUNCTION\_BY\_IMPLICATION} & \textbf{1} & \textbf{4.17\%} & \textbf{1} & \textbf{5.88\%} \\
\textbf{\quad Transformations combined} & \textbf{1} & \textbf{100.00\%} & \textbf{1} & \textbf{100.00\%} \\
\end{longtable}

\clearpage

\begin{longtable}{lrrrr}
\caption{Hand-picked errors and transformations for linguistic operator(s) IF\_IF}\label{tab:IF_IF-errors-trafos}\\
\toprule
Hand-picked mistake / Transformations & \# & \% & \#dst & \%dst \\
\midrule
\endfirsthead
\toprule
Hand-picked mistake / Transformations & \# & \% & \#dst & \%dst \\
\midrule
\endhead
\midrule
\multicolumn{5}{r}{Continued on next page} \\
\endfoot
\bottomrule
\endlastfoot
\midrule
\textbf{BUGGY\_REPLACE\_CONJUNCTION\_BY\_IMPLICATION} & \textbf{40} & \textbf{29.63\%} & \textbf{2} & \textbf{4.00\%} \\
\textbf{\quad Transformations combined} & \textbf{40} & \textbf{100.00\%} & \textbf{2} & \textbf{100.00\%} \\
\quad T1: $Y_{1} \land Y_{2}$ $\rightsquigarrow$ $Y_{2} \rightarrow Y_{1}$ & 40 & 100.00\% & 2 & 100.00\% \\
\quad \quad T16: $\left(Y_{1} \rightarrow Y_{2}\right) \land Y_{3}$ $\rightsquigarrow$ $Y_{3} \rightarrow \left(Y_{1} \rightarrow Y_{2}\right)$ & 13 & 32.50\% & 1 & 50.00\% \\
\midrule
\textbf{BUGGY\_REPLACE\_EQUIVALENCE\_BY\_IMPLICATION} & \textbf{10} & \textbf{7.41\%} & \textbf{2} & \textbf{4.00\%} \\
\textbf{\quad Transformations combined} & \textbf{9} & \textbf{90.00\%} & \textbf{1} & \textbf{50.00\%} \\
\quad $\lceil$ $=$ T8: $Y_{1} \rightarrow \left(Y_{2} \leftrightarrow Y_{3}\right)$ $\rightsquigarrow$ $Y_{1} \rightarrow \left(Y_{2} \rightarrow Y_{3}\right)$ & 9 & 90.00\% & 1 & 50.00\% \\
\quad $\lfloor$ $=$ T17: $Y_{1} \leftrightarrow Y_{2}$ $\rightsquigarrow$ $Y_{1} \rightarrow Y_{2}$ & 9 & 90.00\% & 1 & 50.00\% \\
\end{longtable}

\clearpage

\begin{longtable}{lrrrr}
\caption{Hand-picked errors and transformations for linguistic operator(s) IF\_NEITHER\_NOR\_NOT}\label{tab:IF_NEITHER_NOR_NOT-errors-trafos}\\
\toprule
Hand-picked mistake / Transformations & \# & \% & \#dst & \%dst \\
\midrule
\endfirsthead
\toprule
Hand-picked mistake / Transformations & \# & \% & \#dst & \%dst \\
\midrule
\endhead
\midrule
\multicolumn{5}{r}{Continued on next page} \\
\endfoot
\bottomrule
\endlastfoot
\midrule
\textbf{BUGGY\_REPLACE\_DISJUNCTION\_BY\_CONJUNCTION} & \textbf{73} & \textbf{20.05\%} & \textbf{3} & \textbf{3.16\%} \\
\textbf{\quad Transformations combined} & \textbf{73} & \textbf{100.00\%} & \textbf{3} & \textbf{100.00\%} \\
\quad $\lceil$ $=$ T1: $\lnot \left(Y_{1} \land Y_{2}\right) \rightarrow \lnot Y_{3}$ $\rightsquigarrow$ $\left(\lnot Y_{1} \land \lnot Y_{2}\right) \rightarrow \lnot Y_{3}$ & 65 & 89.04\% & 2 & 66.67\% \\
\quad $|$ $=$ T2: $\lnot \left(Y_{1} \land Y_{2}\right)$ $\rightsquigarrow$ $\lnot Y_{1} \land \lnot Y_{2}$ & 65 & 89.04\% & 2 & 66.67\% \\
\quad $|$ $=$ T3: $\lnot \left(Y_{1} \land Y_{2}\right) \rightarrow Y_{3}$ $\rightsquigarrow$ $\left(\lnot Y_{1} \land \lnot Y_{2}\right) \rightarrow Y_{3}$ & 65 & 89.04\% & 2 & 66.67\% \\
\quad $\lfloor$ $=$ T28: $\lnot \left(Y_{1} \land \ldots \land Y_{n}\right)$ $\rightsquigarrow$ $\lnot Y_{1} \land \ldots \land \lnot Y_{n}$ & 65 & 89.04\% & 2 & 66.67\% \\
\quad T20: $\left(\lnot Y_{1} \lor \lnot Y_{2}\right) \rightarrow \lnot Y_{3}$ $\rightsquigarrow$ $\left(\lnot Y_{1} \land \lnot Y_{2}\right) \rightarrow \lnot Y_{3}$ & 8 & 10.96\% & 1 & 33.33\% \\
\midrule
\textbf{BUGGY\_ADD\_NEGATION} & \textbf{61} & \textbf{16.76\%} & \textbf{4} & \textbf{4.21\%} \\
\end{longtable}

\clearpage

\begin{longtable}{lrrrr}
\caption{Hand-picked errors and transformations for linguistic operator(s) IF\_NEITHER\_NOR}\label{tab:IF_NEITHER_NOR-errors-trafos}\\
\toprule
Hand-picked mistake / Transformations & \# & \% & \#dst & \%dst \\
\midrule
\endfirsthead
\toprule
Hand-picked mistake / Transformations & \# & \% & \#dst & \%dst \\
\midrule
\endhead
\midrule
\multicolumn{5}{r}{Continued on next page} \\
\endfoot
\bottomrule
\endlastfoot
\midrule
\textbf{BUGGY\_REPLACE\_DISJUNCTION\_BY\_CONJUNCTION} & \textbf{37} & \textbf{45.12\%} & \textbf{2} & \textbf{6.25\%} \\
\textbf{\quad Transformations combined} & \textbf{37} & \textbf{100.00\%} & \textbf{2} & \textbf{100.00\%} \\
\quad T1: $\lnot \left(Y_{1} \land Y_{2}\right) \rightarrow Y_{3}$ $\rightsquigarrow$ $\left(\lnot Y_{1} \land \lnot Y_{2}\right) \rightarrow Y_{3}$ & 28 & 75.68\% & 1 & 50.00\% \\
\quad T2: $\left(\lnot Y_{1} \lor \lnot Y_{2}\right) \rightarrow Y_{3}$ $\rightsquigarrow$ $\left(\lnot Y_{1} \land \lnot Y_{2}\right) \rightarrow Y_{3}$ & 9 & 24.32\% & 1 & 50.00\% \\
\midrule
\textbf{BUGGY\_REPLACE\_IMPLICATION\_BY\_REVERSED\_IMPLICATION} & \textbf{5} & \textbf{6.10\%} & \textbf{2} & \textbf{6.25\%} \\
\textbf{\quad Transformations combined} & \textbf{5} & \textbf{100.00\%} & \textbf{2} & \textbf{100.00\%} \\
\quad T4: $Y_{1} \rightarrow \left(\lnot Y_{2} \land \lnot Y_{3}\right)$ $\rightsquigarrow$ $\left(\lnot Y_{2} \land \lnot Y_{3}\right) \rightarrow Y_{1}$ & 4 & 80.00\% & 1 & 50.00\% \\
\quad T14: $Y_{1} \rightarrow \lnot \left(Y_{2} \lor Y_{3}\right)$ $\rightsquigarrow$ $\left(\lnot Y_{2} \land \lnot Y_{3}\right) \rightarrow Y_{1}$ & 1 & 20.00\% & 1 & 50.00\% \\
\end{longtable}

\clearpage

\begin{longtable}{lrrrr}
\caption{Hand-picked errors and transformations for linguistic operator(s) IF\_NOT\_AND}\label{tab:IF_NOT_AND-errors-trafos}\\
\toprule
Hand-picked mistake / Transformations & \# & \% & \#dst & \%dst \\
\midrule
\endfirsthead
\toprule
Hand-picked mistake / Transformations & \# & \% & \#dst & \%dst \\
\midrule
\endhead
\midrule
\multicolumn{5}{r}{Continued on next page} \\
\endfoot
\bottomrule
\endlastfoot
\midrule
\textbf{BUGGY\_ADD\_NEGATION} & \textbf{16} & \textbf{39.02\%} & \textbf{2} & \textbf{12.50\%} \\
\textbf{\quad Transformations combined} & \textbf{16} & \textbf{100.00\%} & \textbf{2} & \textbf{100.00\%} \\
\quad T1: $\left(Y_{1} \land Y_{2}\right) \rightarrow Y_{3}$ $\rightsquigarrow$ $\left(Y_{2} \land \lnot Y_{1}\right) \rightarrow Y_{3}$ & 15 & 93.75\% & 1 & 50.00\% \\
\quad T11: $Y_{1} \lor \lnot \left(Y_{2} \land Y_{3}\right)$ $\rightsquigarrow$ $\left(Y_{3} \land \lnot Y_{2}\right) \rightarrow Y_{1}$ & 1 & 6.25\% & 1 & 50.00\% \\
\midrule
\textbf{BUGGY\_REMOVE\_NEGATION} & \textbf{6} & \textbf{14.63\%} & \textbf{2} & \textbf{12.50\%} \\
\textbf{\quad Transformations combined} & \textbf{6} & \textbf{100.00\%} & \textbf{2} & \textbf{100.00\%} \\
\quad T2: $\left(\lnot Y_{1} \land Y_{2}\right) \rightarrow Y_{3}$ $\rightsquigarrow$ $\left(Y_{1} \land Y_{2}\right) \rightarrow Y_{3}$ & 5 & 83.33\% & 1 & 50.00\% \\
\quad T7: $\lnot \left(Y_{1} \lor Y_{2}\right)$ $\rightsquigarrow$ $Y_{1} \land \lnot Y_{2}$ & 1 & 16.67\% & 1 & 50.00\% \\
\end{longtable}

\clearpage

\begin{longtable}{lrrrr}
\caption{Hand-picked errors and transformations for linguistic operator(s) IF\_NOT}\label{tab:IF_NOT-errors-trafos}\\
\toprule
Hand-picked mistake / Transformations & \# & \% & \#dst & \%dst \\
\midrule
\endfirsthead
\toprule
Hand-picked mistake / Transformations & \# & \% & \#dst & \%dst \\
\midrule
\endhead
\midrule
\multicolumn{5}{r}{Continued on next page} \\
\endfoot
\bottomrule
\endlastfoot
\midrule
\textbf{BUGGY\_REPLACE\_IMPLICATION\_BY\_REVERSED\_IMPLICATION} & \textbf{137} & \textbf{47.24\%} & \textbf{1} & \textbf{4.35\%} \\
\textbf{\quad Transformations combined} & \textbf{137} & \textbf{100.00\%} & \textbf{1} & \textbf{100.00\%} \\
\quad T1: $\lnot Y_{1} \rightarrow Y_{2}$ $\rightsquigarrow$ $Y_{2} \rightarrow \lnot Y_{1}$ & 137 & 100.00\% & 1 & 100.00\% \\
\midrule
\textbf{BUGGY\_REPLACE\_EQUIVALENCE\_BY\_REVERSED\_IMPLICATION} & \textbf{54} & \textbf{18.62\%} & \textbf{2} & \textbf{8.70\%} \\
\textbf{\quad Transformations combined} & \textbf{54} & \textbf{100.00\%} & \textbf{2} & \textbf{100.00\%} \\
\quad T2: $\lnot Y_{1} \leftrightarrow Y_{2}$ $\rightsquigarrow$ $Y_{2} \rightarrow \lnot Y_{1}$ & 49 & 90.74\% & 1 & 50.00\% \\
\quad T8: $Y_{1} \leftrightarrow \lnot Y_{2}$ $\rightsquigarrow$ $Y_{2} \rightarrow \lnot Y_{1}$ & 5 & 9.26\% & 1 & 50.00\% \\
\midrule
\textbf{BUGGY\_REPLACE\_EQUIVALENCE\_BY\_IMPLICATION} & \textbf{20} & \textbf{6.90\%} & \textbf{2} & \textbf{8.70\%} \\
\end{longtable}

\clearpage

\begin{longtable}{lrrrr}
\caption{Hand-picked errors and transformations for linguistic operator(s) IF\_OR\_NEUTRAL}\label{tab:IF_OR_NEUTRAL-errors-trafos}\\
\toprule
Hand-picked mistake / Transformations & \# & \% & \#dst & \%dst \\
\midrule
\endfirsthead
\toprule
Hand-picked mistake / Transformations & \# & \% & \#dst & \%dst \\
\midrule
\endhead
\midrule
\multicolumn{5}{r}{Continued on next page} \\
\endfoot
\bottomrule
\endlastfoot
\midrule
\textbf{BUGGY\_REPLACE\_CONJUNCTION\_BY\_DISJUNCTION} & \textbf{6} & \textbf{17.65\%} & \textbf{2} & \textbf{9.09\%} \\
\textbf{\quad Transformations combined} & \textbf{6} & \textbf{100.00\%} & \textbf{2} & \textbf{100.00\%} \\
\quad T1: $Y_{1} \rightarrow \left(Y_{2} \land Y_{3}\right)$ $\rightsquigarrow$ $Y_{1} \rightarrow \left(Y_{2} \lor Y_{3}\right)$ & 6 & 100.00\% & 2 & 100.00\% \\
\midrule
\textbf{BUGGY\_REMOVE\_NEGATION} & \textbf{5} & \textbf{14.71\%} & \textbf{2} & \textbf{9.09\%} \\
\textbf{\quad Transformations combined} & \textbf{5} & \textbf{100.00\%} & \textbf{2} & \textbf{100.00\%} \\
\quad T2: $Y_{1} \rightarrow \left(Y_{2} \lor \lnot Y_{3}\right)$ $\rightsquigarrow$ $Y_{1} \rightarrow \left(Y_{2} \lor Y_{3}\right)$ & 4 & 80.00\% & 1 & 50.00\% \\
\quad T5: $\lnot Y_{1} \rightarrow Y_{2}$ $\rightsquigarrow$ $Y_{1} \rightarrow Y_{2}$ & 1 & 20.00\% & 1 & 50.00\% \\
\midrule
\textbf{BUGGY\_REPLACE\_IMPLICATION\_BY\_REVERSED\_IMPLICATION} & \textbf{4} & \textbf{11.76\%} & \textbf{1} & \textbf{4.55\%} \\
\end{longtable}

\clearpage

\begin{longtable}{lrrrr}
\caption{Hand-picked errors and transformations for linguistic operator(s) IF\_OR}\label{tab:IF_OR-errors-trafos}\\
\toprule
Hand-picked mistake / Transformations & \# & \% & \#dst & \%dst \\
\midrule
\endfirsthead
\toprule
Hand-picked mistake / Transformations & \# & \% & \#dst & \%dst \\
\midrule
\endhead
\midrule
\multicolumn{5}{r}{Continued on next page} \\
\endfoot
\bottomrule
\endlastfoot
\midrule
\textbf{BUGGY\_REPLACE\_IMPLICATION\_BY\_REVERSED\_IMPLICATION} & \textbf{18} & \textbf{52.94\%} & \textbf{4} & \textbf{23.53\%} \\
\textbf{\quad Transformations combined} & \textbf{18} & \textbf{100.00\%} & \textbf{4} & \textbf{100.00\%} \\
\quad $\lceil$ $=$ T1: $\left(Y_{1} \lor Y_{2}\right) \rightarrow Y_{3}$ $\rightsquigarrow$ $Y_{3} \rightarrow \left(Y_{1} \lor Y_{2}\right)$ & 18 & 100.00\% & 4 & 100.00\% \\
\quad $\lfloor$ $=$ T9: $Y_{1} \rightarrow Y_{2}$ $\rightsquigarrow$ $Y_{2} \rightarrow Y_{1}$ & 18 & 100.00\% & 4 & 100.00\% \\
\midrule
\textbf{BUGGY\_REPLACE\_CONJUNCTION\_BY\_DISJUNCTION} & \textbf{3} & \textbf{8.82\%} & \textbf{2} & \textbf{11.76\%} \\
\textbf{\quad Transformations combined} & \textbf{3} & \textbf{100.00\%} & \textbf{2} & \textbf{100.00\%} \\
\quad T3: $Y_{1} \rightarrow \left(Y_{2} \land Y_{3}\right)$ $\rightsquigarrow$ $Y_{1} \rightarrow \left(Y_{2} \lor Y_{3}\right)$ & 3 & 100.00\% & 2 & 100.00\% \\
\midrule
\textbf{BUGGY\_REPLACE\_EQUIVALENCE\_BY\_REVERSED\_IMPLICATION} & \textbf{3} & \textbf{8.82\%} & \textbf{2} & \textbf{11.76\%} \\
\end{longtable}

\clearpage

\begin{longtable}{lrrrr}
\caption{Hand-picked errors and transformations for linguistic operator(s) IF}\label{tab:IF-errors-trafos}\\
\toprule
Hand-picked mistake / Transformations & \# & \% & \#dst & \%dst \\
\midrule
\endfirsthead
\toprule
Hand-picked mistake / Transformations & \# & \% & \#dst & \%dst \\
\midrule
\endhead
\midrule
\multicolumn{5}{r}{Continued on next page} \\
\endfoot
\bottomrule
\endlastfoot
\midrule
\textbf{BUGGY\_REPLACE\_IMPLICATION\_BY\_REVERSED\_IMPLICATION} & \textbf{61} & \textbf{58.65\%} & \textbf{4} & \textbf{19.05\%} \\
\textbf{\quad Transformations combined} & \textbf{61} & \textbf{100.00\%} & \textbf{4} & \textbf{100.00\%} \\
\quad T1: $Y_{1} \rightarrow Y_{2}$ $\rightsquigarrow$ $Y_{2} \rightarrow Y_{1}$ & 61 & 100.00\% & 4 & 100.00\% \\
\midrule
\textbf{BUGGY\_REPLACE\_EQUIVALENCE\_BY\_IMPLICATION} & \textbf{20} & \textbf{19.23\%} & \textbf{4} & \textbf{19.05\%} \\
\textbf{\quad Transformations combined} & \textbf{20} & \textbf{100.00\%} & \textbf{4} & \textbf{100.00\%} \\
\quad T2: $Y_{1} \leftrightarrow Y_{2}$ $\rightsquigarrow$ $Y_{1} \rightarrow Y_{2}$ & 20 & 100.00\% & 4 & 100.00\% \\
\midrule
\textbf{BUGGY\_REPLACE\_CONJUNCTION\_BY\_IMPLICATION} & \textbf{10} & \textbf{9.62\%} & \textbf{4} & \textbf{19.05\%} \\
\textbf{\quad Transformations combined} & \textbf{10} & \textbf{100.00\%} & \textbf{4} & \textbf{100.00\%} \\
\end{longtable}

\clearpage

\begin{longtable}{lrrrr}
\caption{Hand-picked errors and transformations for linguistic operator(s) NEITHER\_NOR}\label{tab:NEITHER_NOR-errors-trafos}\\
\toprule
Hand-picked mistake / Transformations & \# & \% & \#dst & \%dst \\
\midrule
\endfirsthead
\toprule
Hand-picked mistake / Transformations & \# & \% & \#dst & \%dst \\
\midrule
\endhead
\midrule
\multicolumn{5}{r}{Continued on next page} \\
\endfoot
\bottomrule
\endlastfoot
\midrule
\textbf{BUGGY\_REPLACE\_DISJUNCTION\_BY\_CONJUNCTION} & \textbf{25} & \textbf{37.31\%} & \textbf{7} & \textbf{23.33\%} \\
\textbf{\quad Transformations combined} & \textbf{25} & \textbf{100.00\%} & \textbf{7} & \textbf{100.00\%} \\
\quad T1: $\lnot \left(Y_{1} \land \ldots \land Y_{n}\right)$ $\rightsquigarrow$ $\lnot Y_{1} \land \ldots \land \lnot Y_{n}$ & 14 & 56.00\% & 2 & 28.57\% \\
\quad $\lceil$ $=$ T2: $Y_{1} \lor \ldots \lor Y_{n}$ $\rightsquigarrow$ $Y_{1} \land \ldots \land Y_{n}$ & 7 & 28.00\% & 2 & 28.57\% \\
\quad $\lfloor$ $=$ T7: $\lnot Y_{1} \lor \ldots \lor \lnot Y_{n}$ $\rightsquigarrow$ $\lnot Y_{1} \land \ldots \land \lnot Y_{n}$ & 7 & 28.00\% & 2 & 28.57\% \\
\quad T11: $\lnot Y_{1} \lor \lnot Y_{2} \lor \lnot Y_{3}$ $\rightsquigarrow$ $\lnot Y_{1} \land \lnot Y_{3} \land \lnot Y_{2}$ & 2 & 8.00\% & 1 & 14.29\% \\
\quad $\lceil$ $=$ T4: $\lnot Y_{1} \lor \lnot Y_{2} \lor \lnot Y_{3}$ $\rightsquigarrow$ $\lnot Y_{1} \land \lnot Y_{2} \land \lnot Y_{3}$ & 1 & 4.00\% & 1 & 14.29\% \\
\quad $\lfloor$ $=$ T9: $Y_{1} \lor \ldots \lor Y_{n} \lor Y_{3}$ $\rightsquigarrow$ $Y_{1} \land \ldots \land Y_{n} \land Y_{3}$ & 1 & 4.00\% & 1 & 14.29\% \\
\end{longtable}

\clearpage

\begin{longtable}{lrrrr}
\caption{Hand-picked errors and transformations for linguistic operator(s) NEUTRAL}\label{tab:NEUTRAL-errors-trafos}\\
\toprule
Hand-picked mistake / Transformations & \# & \% & \#dst & \%dst \\
\midrule
\endfirsthead
\toprule
Hand-picked mistake / Transformations & \# & \% & \#dst & \%dst \\
\midrule
\endhead
\midrule
\multicolumn{5}{r}{Continued on next page} \\
\endfoot
\bottomrule
\endlastfoot
\midrule
\textbf{BUGGY\_REMOVE\_NEGATION} & \textbf{52} & \textbf{100.00\%} & \textbf{2} & \textbf{100.00\%} \\
\textbf{\quad Transformations combined} & \textbf{52} & \textbf{100.00\%} & \textbf{2} & \textbf{100.00\%} \\
\quad T1: $\lnot Y_{1}$ $\rightsquigarrow$ $Y_{1}$ & 52 & 100.00\% & 2 & 100.00\% \\
\midrule
\textbf{Unexplained by ALL hand errors} & \textbf{0} & \textbf{0.00\%} & \textbf{0} & \textbf{0.00\%} \\
\textbf{\quad Transformations combined} & \textbf{0} & \textbf{0.00\%} & \textbf{0} & \textbf{0.00\%} \\
\midrule
Whole file & 52 & 100.00\% & 2 & 100.00\% \\
\textbf{ANY hand error explained} & \textbf{52} & \textbf{100.00\%} & \textbf{2} & \textbf{100.00\%} \\
\textbf{ANY transformation explained} & \textbf{52} & \textbf{100.00\%} & \textbf{2} & \textbf{100.00\%} \\
Only explained by hand error & 0 & 0.00\% & 0 & 0.00\% \\
Only explained by transformation & 0 & 0.00\% & 0 & 0.00\% \\
\end{longtable}

\clearpage

\begin{longtable}{lrrrr}
\caption{Hand-picked errors and transformations for linguistic operator(s) NONE\_OF}\label{tab:NONE_OF-errors-trafos}\\
\toprule
Hand-picked mistake / Transformations & \# & \% & \#dst & \%dst \\
\midrule
\endfirsthead
\toprule
Hand-picked mistake / Transformations & \# & \% & \#dst & \%dst \\
\midrule
\endhead
\midrule
\multicolumn{5}{r}{Continued on next page} \\
\endfoot
\bottomrule
\endlastfoot
\midrule
\textbf{BUGGY\_REPLACE\_DISJUNCTION\_BY\_CONJUNCTION} & \textbf{36} & \textbf{78.26\%} & \textbf{7} & \textbf{46.67\%} \\
\textbf{\quad Transformations combined} & \textbf{36} & \textbf{100.00\%} & \textbf{7} & \textbf{100.00\%} \\
\quad T1: $\lnot \left(Y_{1} \land \ldots \land Y_{n}\right)$ $\rightsquigarrow$ $\lnot Y_{1} \land \ldots \land \lnot Y_{n}$ & 34 & 94.44\% & 5 & 71.43\% \\
\quad T7: $Y_{1} \lor \ldots \lor Y_{n}$ $\rightsquigarrow$ $Y_{1} \land \ldots \land Y_{n}$ & 2 & 5.56\% & 2 & 28.57\% \\
\midrule
\textbf{BUGGY\_REPLACE\_NOR\_BY\_CONJUNCTION} & \textbf{4} & \textbf{8.70\%} & \textbf{2} & \textbf{13.33\%} \\
\textbf{\quad Transformations combined} & \textbf{4} & \textbf{100.00\%} & \textbf{2} & \textbf{100.00\%} \\
\quad T2: $Y_{1} \land \ldots \land Y_{n}$ $\rightsquigarrow$ $\lnot Y_{1} \land \ldots \land \lnot Y_{n}$ & 4 & 100.00\% & 2 & 100.00\% \\
\midrule
\textbf{BUGGY\_REPLACE\_NAND\_BY\_CONJUNCTION} & \textbf{2} & \textbf{4.35\%} & \textbf{2} & \textbf{13.33\%} \\
\end{longtable}

\clearpage

\begin{longtable}{lrrrr}
\caption{Hand-picked errors and transformations for linguistic operator(s) NOT\_ALL\_OF}\label{tab:NOT_ALL_OF-errors-trafos}\\
\toprule
Hand-picked mistake / Transformations & \# & \% & \#dst & \%dst \\
\midrule
\endfirsthead
\toprule
Hand-picked mistake / Transformations & \# & \% & \#dst & \%dst \\
\midrule
\endhead
\midrule
\multicolumn{5}{r}{Continued on next page} \\
\endfoot
\bottomrule
\endlastfoot
\midrule
\textbf{BUGGY\_PULL\_NEGATION\_CONJUNCTION} & \textbf{67} & \textbf{12.48\%} & \textbf{13} & \textbf{7.78\%} \\
\textbf{\quad Transformations combined} & \textbf{67} & \textbf{100.00\%} & \textbf{13} & \textbf{100.00\%} \\
\quad $\lceil$ $=$ T3: $\lnot Y_{1} \land \lnot Y_{2}$ $\rightsquigarrow$ $\lnot \left(Y_{1} \land Y_{2}\right)$ & 61 & 91.04\% & 12 & 92.31\% \\
\quad $\lfloor$ $=$ T4: $\lnot Y_{1} \land \ldots \land \lnot Y_{n}$ $\rightsquigarrow$ $\lnot \left(Y_{1} \land \ldots \land Y_{n}\right)$ & 61 & 91.04\% & 12 & 92.31\% \\
\quad T17: $\lnot Y_{1} \land \ldots \land \lnot Y_{n}$ $\rightsquigarrow$ $\lnot \left(Y_{1} \land \ldots \land Y_{n}\right)$ & 6 & 8.96\% & 1 & 7.69\% \\
\midrule
\textbf{BUGGY\_REPLACE\_XOR\_BY\_NAND} & \textbf{44} & \textbf{8.19\%} & \textbf{5} & \textbf{2.99\%} \\
\textbf{\quad Transformations combined} & \textbf{43} & \textbf{97.73\%} & \textbf{4} & \textbf{80.00\%} \\
\quad T7: $\lnot \left(Y_{1} \leftrightarrow Y_{2}\right)$ $\rightsquigarrow$ $\lnot \left(Y_{1} \land Y_{2}\right)$ & 24 & 54.55\% & 2 & 40.00\% \\
\end{longtable}

\clearpage

\begin{longtable}{lrrrr}
\caption{Hand-picked errors and transformations for linguistic operator(s) NOT}\label{tab:NOT-errors-trafos}\\
\toprule
Hand-picked mistake / Transformations & \# & \% & \#dst & \%dst \\
\midrule
\endfirsthead
\toprule
Hand-picked mistake / Transformations & \# & \% & \#dst & \%dst \\
\midrule
\endhead
\midrule
\multicolumn{5}{r}{Continued on next page} \\
\endfoot
\bottomrule
\endlastfoot
\midrule
\textbf{BUGGY\_ADD\_NEGATION} & \textbf{1} & \textbf{100.00\%} & \textbf{1} & \textbf{100.00\%} \\
\textbf{\quad Transformations combined} & \textbf{1} & \textbf{100.00\%} & \textbf{1} & \textbf{100.00\%} \\
\quad T1: $Y_{1}$ $\rightsquigarrow$ $\lnot Y_{1}$ & 1 & 100.00\% & 1 & 100.00\% \\
\midrule
\textbf{Unexplained by ALL hand errors} & \textbf{0} & \textbf{0.00\%} & \textbf{0} & \textbf{0.00\%} \\
\textbf{\quad Transformations combined} & \textbf{0} & \textbf{0.00\%} & \textbf{0} & \textbf{0.00\%} \\
\midrule
Whole file & 1 & 100.00\% & 1 & 100.00\% \\
\textbf{ANY hand error explained} & \textbf{1} & \textbf{100.00\%} & \textbf{1} & \textbf{100.00\%} \\
\textbf{ANY transformation explained} & \textbf{1} & \textbf{100.00\%} & \textbf{1} & \textbf{100.00\%} \\
Only explained by hand error & 0 & 0.00\% & 0 & 0.00\% \\
Only explained by transformation & 0 & 0.00\% & 0 & 0.00\% \\
\end{longtable}

\clearpage

\begin{longtable}{lrrrr}
\caption{Hand-picked errors and transformations for linguistic operator(s) ONLY\_IF\_AND}\label{tab:ONLY_IF_AND-errors-trafos}\\
\toprule
Hand-picked mistake / Transformations & \# & \% & \#dst & \%dst \\
\midrule
\endfirsthead
\toprule
Hand-picked mistake / Transformations & \# & \% & \#dst & \%dst \\
\midrule
\endhead
\midrule
\multicolumn{5}{r}{Continued on next page} \\
\endfoot
\bottomrule
\endlastfoot
\midrule
\textbf{BUGGY\_REPLACE\_IMPLICATION\_BY\_REVERSED\_IMPLICATION} & \textbf{446} & \textbf{50.06\%} & \textbf{5} & \textbf{5.75\%} \\
\textbf{\quad Transformations combined} & \textbf{443} & \textbf{99.33\%} & \textbf{3} & \textbf{60.00\%} \\
\quad $\lceil$ $=$ T1: $\left(Y_{1} \land \ldots \land Y_{n}\right) \rightarrow Y_{3}$ $\rightsquigarrow$ $Y_{3} \rightarrow \left(Y_{1} \land \ldots \land Y_{n}\right)$ & 443 & 99.33\% & 3 & 60.00\% \\
\quad $|$ $=$ T2: $\left(Y_{1} \land Y_{2}\right) \rightarrow Y_{3}$ $\rightsquigarrow$ $Y_{3} \rightarrow \left(Y_{1} \land Y_{2}\right)$ & 443 & 99.33\% & 3 & 60.00\% \\
\quad $\lfloor$ $=$ T3: $Y_{1} \rightarrow Y_{2}$ $\rightsquigarrow$ $Y_{2} \rightarrow Y_{1}$ & 443 & 99.33\% & 3 & 60.00\% \\
\midrule
\textbf{BUGGY\_REPLACE\_EQUIVALENCE\_BY\_REVERSED\_IMPLICATION} & \textbf{202} & \textbf{22.67\%} & \textbf{4} & \textbf{4.60\%} \\
\textbf{\quad Transformations combined} & \textbf{201} & \textbf{99.50\%} & \textbf{3} & \textbf{75.00\%} \\
\quad $\lceil$ $=$ T4: $\left(Y_{1} \land \ldots \land Y_{n}\right) \leftrightarrow Y_{3}$ $\rightsquigarrow$ $Y_{3} \rightarrow \left(Y_{1} \land \ldots \land Y_{n}\right)$ & 201 & 99.50\% & 3 & 75.00\% \\
\end{longtable}

\clearpage

\begin{longtable}{lrrrr}
\caption{Hand-picked errors and transformations for linguistic operator(s) ONLY\_IF\_AT\_LEAST}\label{tab:ONLY_IF_AT_LEAST-errors-trafos}\\
\toprule
Hand-picked mistake / Transformations & \# & \% & \#dst & \%dst \\
\midrule
\endfirsthead
\toprule
Hand-picked mistake / Transformations & \# & \% & \#dst & \%dst \\
\midrule
\endhead
\midrule
\multicolumn{5}{r}{Continued on next page} \\
\endfoot
\bottomrule
\endlastfoot
\midrule
\textbf{BUGGY\_REPLACE\_IMPLICATION\_BY\_REVERSED\_IMPLICATION} & \textbf{277} & \textbf{45.26\%} & \textbf{20} & \textbf{10.10\%} \\
\textbf{\quad Transformations combined} & \textbf{274} & \textbf{98.92\%} & \textbf{18} & \textbf{90.00\%} \\
\quad T3: $\left(Y_{1} \lor Y_{2}\right) \rightarrow Y_{3}$ $\rightsquigarrow$ $Y_{3} \rightarrow \left(Y_{1} \lor Y_{2}\right)$ & 239 & 86.28\% & 4 & 20.00\% \\
\quad T7: $\left(Y_{1} \lor Y_{2} \lor Y_{3}\right) \rightarrow Y_{4}$ $\rightsquigarrow$ $Y_{4} \rightarrow \left(Y_{1} \lor Y_{2} \lor Y_{3}\right)$ & 35 & 12.64\% & 14 & 70.00\% \\
\midrule
\textbf{BUGGY\_REPLACE\_EQUIVALENCE\_BY\_REVERSED\_IMPLICATION} & \textbf{33} & \textbf{5.39\%} & \textbf{4} & \textbf{2.02\%} \\
\textbf{\quad Transformations combined} & \textbf{33} & \textbf{100.00\%} & \textbf{4} & \textbf{100.00\%} \\
\quad T1: $Y_{1} \leftrightarrow Y_{2}$ $\rightsquigarrow$ $Y_{2} \rightarrow Y_{1}$ & 33 & 100.00\% & 4 & 100.00\% \\
\midrule
\textbf{BUGGY\_REPLACE\_CONJUNCTION\_BY\_IMPLICATION} & \textbf{29} & \textbf{4.74\%} & \textbf{17} & \textbf{8.59\%} \\
\end{longtable}

\clearpage

\begin{longtable}{lrrrr}
\caption{Hand-picked errors and transformations for linguistic operator(s) ONLY\_IF\_NEITHER\_NOR}\label{tab:ONLY_IF_NEITHER_NOR-errors-trafos}\\
\toprule
Hand-picked mistake / Transformations & \# & \% & \#dst & \%dst \\
\midrule
\endfirsthead
\toprule
Hand-picked mistake / Transformations & \# & \% & \#dst & \%dst \\
\midrule
\endhead
\midrule
\multicolumn{5}{r}{Continued on next page} \\
\endfoot
\bottomrule
\endlastfoot
\midrule
\textbf{BUGGY\_REPLACE\_IMPLICATION\_BY\_REVERSED\_IMPLICATION} & \textbf{138} & \textbf{28.34\%} & \textbf{33} & \textbf{15.21\%} \\
\textbf{\quad Transformations combined} & \textbf{138} & \textbf{100.00\%} & \textbf{33} & \textbf{100.00\%} \\
\quad T1: $Y_{1} \rightarrow Y_{2}$ $\rightsquigarrow$ $Y_{2} \rightarrow Y_{1}$ & 117 & 84.78\% & 20 & 60.61\% \\
\quad T4: $Y_{1} \rightarrow \lnot \left(Y_{2} \lor Y_{3}\right)$ $\rightsquigarrow$ $\left(\lnot Y_{2} \land \lnot Y_{3}\right) \rightarrow Y_{1}$ & 21 & 15.22\% & 13 & 39.39\% \\
\midrule
\textbf{BUGGY\_REPLACE\_DISJUNCTION\_BY\_CONJUNCTION} & \textbf{47} & \textbf{9.65\%} & \textbf{17} & \textbf{7.83\%} \\
\textbf{\quad Transformations combined} & \textbf{47} & \textbf{100.00\%} & \textbf{17} & \textbf{100.00\%} \\
\quad $\lceil$ $=$ T2: $\lnot \left(Y_{1} \land Y_{2}\right) \rightarrow Y_{3}$ $\rightsquigarrow$ $\left(\lnot Y_{1} \land \lnot Y_{2}\right) \rightarrow Y_{3}$ & 29 & 61.70\% & 10 & 58.82\% \\
\quad $\lfloor$ $=$ T3: $\lnot \left(Y_{1} \land \ldots \land Y_{n}\right)$ $\rightsquigarrow$ $\lnot Y_{1} \land \ldots \land \lnot Y_{n}$ & 29 & 61.70\% & 10 & 58.82\% \\
\end{longtable}

\clearpage

\begin{longtable}{lrrrr}
\caption{Hand-picked errors and transformations for linguistic operator(s) ONLY\_IF\_NOT\_AND\_NOT\_NOT}\label{tab:ONLY_IF_NOT_AND_NOT_NOT-errors-trafos}\\
\toprule
Hand-picked mistake / Transformations & \# & \% & \#dst & \%dst \\
\midrule
\endfirsthead
\toprule
Hand-picked mistake / Transformations & \# & \% & \#dst & \%dst \\
\midrule
\endhead
\midrule
\multicolumn{5}{r}{Continued on next page} \\
\endfoot
\bottomrule
\endlastfoot
\midrule
\textbf{BUGGY\_REPLACE\_IMPLICATION\_BY\_REVERSED\_IMPLICATION} & \textbf{19} & \textbf{32.76\%} & \textbf{2} & \textbf{11.11\%} \\
\textbf{\quad Transformations combined} & \textbf{19} & \textbf{100.00\%} & \textbf{2} & \textbf{100.00\%} \\
\quad T1: $Y_{1} \rightarrow Y_{2}$ $\rightsquigarrow$ $Y_{2} \rightarrow Y_{1}$ & 11 & 57.89\% & 1 & 50.00\% \\
\quad T7: $\lnot \left(Y_{1} \lor Y_{2}\right) \rightarrow \lnot Y_{3}$ $\rightsquigarrow$ $\lnot Y_{3} \rightarrow \left(\lnot Y_{1} \land \lnot Y_{2}\right)$ & 8 & 42.11\% & 1 & 50.00\% \\
\midrule
\textbf{BUGGY\_ADD\_NEGATION} & \textbf{3} & \textbf{5.17\%} & \textbf{2} & \textbf{11.11\%} \\
\textbf{\quad Transformations combined} & \textbf{3} & \textbf{100.00\%} & \textbf{2} & \textbf{100.00\%} \\
\quad T10: $Y_{1} \rightarrow \lnot \left(Y_{2} \lor Y_{3}\right)$ $\rightsquigarrow$ $\lnot Y_{1} \rightarrow \left(\lnot Y_{2} \land \lnot Y_{3}\right)$ & 2 & 66.67\% & 1 & 50.00\% \\
\quad T16: $Y_{1} \rightarrow Y_{2}$ $\rightsquigarrow$ $\lnot Y_{1} \rightarrow Y_{2}$ & 1 & 33.33\% & 1 & 50.00\% \\
\end{longtable}

\clearpage

\begin{longtable}{lrrrr}
\caption{Hand-picked errors and transformations for linguistic operator(s) ONLY\_IF\_NOT\_OR\_NOT\_NOT}\label{tab:ONLY_IF_NOT_OR_NOT_NOT-errors-trafos}\\
\toprule
Hand-picked mistake / Transformations & \# & \% & \#dst & \%dst \\
\midrule
\endfirsthead
\toprule
Hand-picked mistake / Transformations & \# & \% & \#dst & \%dst \\
\midrule
\endhead
\midrule
\multicolumn{5}{r}{Continued on next page} \\
\endfoot
\bottomrule
\endlastfoot
\midrule
\textbf{BUGGY\_REPLACE\_IMPLICATION\_BY\_REVERSED\_IMPLICATION} & \textbf{50} & \textbf{30.12\%} & \textbf{10} & \textbf{12.05\%} \\
\textbf{\quad Transformations combined} & \textbf{50} & \textbf{100.00\%} & \textbf{10} & \textbf{100.00\%} \\
\quad T1: $Y_{1} \rightarrow Y_{2}$ $\rightsquigarrow$ $Y_{2} \rightarrow Y_{1}$ & 44 & 88.00\% & 7 & 70.00\% \\
\quad T11: $\lnot \left(Y_{1} \land Y_{2}\right) \rightarrow \lnot Y_{3}$ $\rightsquigarrow$ $\lnot Y_{3} \rightarrow \left(\lnot Y_{1} \lor \lnot Y_{2}\right)$ & 6 & 12.00\% & 3 & 30.00\% \\
\midrule
\textbf{BUGGY\_ADD\_NEGATION} & \textbf{21} & \textbf{12.65\%} & \textbf{10} & \textbf{12.05\%} \\
\textbf{\quad Transformations combined} & \textbf{21} & \textbf{100.00\%} & \textbf{10} & \textbf{100.00\%} \\
\quad T2: $Y_{1} \rightarrow Y_{2}$ $\rightsquigarrow$ $\lnot Y_{1} \rightarrow Y_{2}$ & 16 & 76.19\% & 6 & 60.00\% \\
\quad T16: $Y_{1} \lor Y_{2}$ $\rightsquigarrow$ $Y_{2} \lor \lnot Y_{1}$ & 3 & 14.29\% & 2 & 20.00\% \\
\end{longtable}

\clearpage

\begin{longtable}{lrrrr}
\caption{Hand-picked errors and transformations for linguistic operator(s) ONLY\_IF}\label{tab:ONLY_IF-errors-trafos}\\
\toprule
Hand-picked mistake / Transformations & \# & \% & \#dst & \%dst \\
\midrule
\endfirsthead
\toprule
Hand-picked mistake / Transformations & \# & \% & \#dst & \%dst \\
\midrule
\endhead
\midrule
\multicolumn{5}{r}{Continued on next page} \\
\endfoot
\bottomrule
\endlastfoot
\midrule
\textbf{BUGGY\_REPLACE\_IMPLICATION\_BY\_REVERSED\_IMPLICATION} & \textbf{164} & \textbf{35.12\%} & \textbf{1} & \textbf{3.70\%} \\
\textbf{\quad Transformations combined} & \textbf{164} & \textbf{100.00\%} & \textbf{1} & \textbf{100.00\%} \\
\quad T1: $Y_{1} \rightarrow Y_{2}$ $\rightsquigarrow$ $Y_{2} \rightarrow Y_{1}$ & 164 & 100.00\% & 1 & 100.00\% \\
\midrule
\textbf{BUGGY\_REPLACE\_EQUIVALENCE\_BY\_IMPLICATION} & \textbf{156} & \textbf{33.40\%} & \textbf{2} & \textbf{7.41\%} \\
\textbf{\quad Transformations combined} & \textbf{152} & \textbf{97.44\%} & \textbf{1} & \textbf{50.00\%} \\
\quad T2: $Y_{1} \leftrightarrow Y_{2}$ $\rightsquigarrow$ $Y_{1} \rightarrow Y_{2}$ & 152 & 97.44\% & 1 & 50.00\% \\
\midrule
\textbf{BUGGY\_REPLACE\_EQUIVALENCE\_BY\_REVERSED\_IMPLICATION} & \textbf{97} & \textbf{20.77\%} & \textbf{3} & \textbf{11.11\%} \\
\textbf{\quad Transformations combined} & \textbf{93} & \textbf{95.88\%} & \textbf{1} & \textbf{33.33\%} \\
\end{longtable}%
 
\clearpage
\subsection{Full Propositional Logic Data Set}
The hand-picked mistakes are from \cite{SchmellenkampLZ23}.
%
 
\clearpage
\subsection{Propositional Logic: Hierarchical Clustering of Transformations}
%
 
\clearpage
\subsection{Modal Logic: Hierarchical Clustering of Transformations}
%
 
\clearpage
\subsection{CTL: Hierarchical Clustering of Transformations}
%
 
\end{document}